\documentclass[fleqn,usenatbib]{mnras}

\usepackage{amssymb}	

\usepackage{newtxtext,newtxmath}
\usepackage{amssymb}	

\usepackage[T1]{fontenc}

\usepackage{graphicx}	
\usepackage{amsmath}	
\usepackage{subcaption}
\usepackage{graphicx}

\title[JWST SMBHs at $z \geq 9$]{Are we surprised to find SMBHs with JWST at $z \geq 9$?}

\author[Schneider et al.]{Raffaella Schneider$^{1,2,3,4}$\thanks{E-mail: raffaella.schneider@uniroma1.it},
Rosa Valiante$^{2,3}$,
Alessandro Trinca$^{1,2,3}$, 
Luca Graziani$^{1,2}$, 
Marta Volonteri$^{5}$,
\newauthor
Roberto Maiolino$^{6,7,8}$
\\
$^{1}$Dipartimento di Fisica, ``Sapienza'' Universit$\grave{a}$ di Roma, Piazzale Aldo Moro 2, 00185 Roma, Italy \\
$^{2}$INAF/Osservatorio Astronomico di Roma, Via di Frascati 33, 00040 Monte Porzio Catone, Italy \\
$^{3}$INFN, Sezione Roma1, Dipartimento di Fisica, ``Sapienza'' Universit$\grave{a}$ di Roma, Piazzale Aldo Moro 2, 00185, Roma, Italy \\
$^{4}$Sapienza School for Advanced Studies, Viale Regina Elena 291, 00161 Roma, Italy\\
$^{5}$Institut d’Astrophysique de Paris, Sorbonne Université, CNRS, UMR 7095, 98 bis bd Arago, 75014 Paris, France \\
$^{6}$Kavli Institute for Cosmology, University of Cambridge, Madingley Road, Cambridge CB3 0HA, UK \\
$^{7}$Cavendish Laboratory, University of Cambridge, 19 JJ Thomson Avenue, Cambridge CB3 0HE, UK\\
$^{8}$Department of Physics and Astronomy, University College London, Gower Street, London WC1E 6BT, UK
}

\date{Accepted 2023. Received 2023; in original form 2023}

\pubyear{2023}

\begin{document}
\label{firstpage}
\pagerange{\pageref{firstpage}--\pageref{lastpage}}
\maketitle

\begin{abstract}
JWST is unveiling for the first time accreting black holes (BHs) with masses of $10^6 - 10^7 M_\odot$ at $z > 4$, with the most distant residing in GNz11 at $z=10.6$. Are we really surprised to find them in the nuclei of $z \simeq 5 - 11$ galaxies? Here we predict the properties of $4 < z < 11$ BHs and their host galaxies considering an Eddington-limited (EL) and a super-Eddington (SE) BH accretion scenario, using the Cosmic Archaeology Tool (CAT) semi-analytical model. We calculate the transmitted spectral energy distribution of CAT synthetic candidates, representative of the BH/galaxy properties of GNz11. We also examine the possibility that the $z = 8.7$ galaxy CEERS-1019 could host an active BH. We find that the luminosity of high-$z$ JWST detected BHs are better reproduced by the SE model, where BHs descend from efficiently growing light and heavy seeds. Conversely, the host galaxy stellar masses are better matched in the EL model, in which all the systems detectable with JWST surveys JADES and CEERS descend from heavy BH seeds.  We support the interpretation that the central point source of GNz11 could be powered by a SE ($\lambda_{\rm Edd} \simeq 2 - 3$) accreting BH with mass $1.5 \times 10^6 M_\odot$, while the emission from CEERS-1019 is dominated by the host galaxy; if it harbours an active BH, we find it to have a mass of  $M_{\rm BH} \simeq 10^7 M_\odot$, and to be accreting at sub-Eddington rates ($\lambda_{\rm Edd} \simeq 0.5$).
\end{abstract}




\section{Introduction}

Understanding how galaxies form and evolve in the first billion years of cosmic evolution is one of the driving
science objectives of the recently launched JWST. 
With its unprecedented sensitivity, 
JWST has already started to revolutionise the field.
JWST/NIRCam observations have enabled to detecting galaxies
at $z \gtrsim 10$ \citep{castellano2022, labbe2022, naidu2022, Adams2023, tacchella2022, tacchella2023, robertson2023}, providing the first observational 
constraints on the UV luminosity function and star formation history in the first 500 Myr of the Universe
\citep{Donnan2023, harikane2022b},
showing a surprising lack of evolution in the number density of bright galaxies at $z \gtrsim 10$.
While some of these sources must be interpreted with caution, given their possible confusion with dusty, star-forming
galaxies at $z \sim 5$ \citep{zavala2023, naidu2022b, arrabalharo2023}, JWST is showing that galaxies in the early Universe are
characterized by a diversity of properties \citep{barrufet2022, whitler2023}, including unexpectedly strong dust 
obscuration at $8 < z < 13$ \citep{rodighiero2023}, massive quiescent galaxies at $3 < z < 5$ \citep{carnall2023}, and even low-mass quenched galaxies out to $z \simeq 7$ \citep{looser2023}.
JWST/NIRSpec observations have spectroscopically confirmed several tens of these systems 
(see \citealt{nakajima2023} and references therein), including galaxies at $10 \leq z \leq 13.2$ \citep{curtislake2022, robertson2023, bunker2023, arrabalharo2023}. Most of these galaxies 
show strong nebular lines, indicative of extreme excitation conditions, with high ionization parameters
and low-metallicities \citep{curti2023, cameron2023, sanders2023a, sanders2023b}. 

Interestingly, the spectra of some of these galaxies reveal the presence of broad permitted 
lines (H$\alpha$, H$\beta$, H$_\gamma$) without a counterpart in the forbidden lines, hence consistent with being emitted from the Broad Line Region (BLR) of 
weak Active Galactic Nuclei (AGNs). Two systems with estimated 
BH masses of $M_{\rm BH} \sim 10^7 M_\odot$ 
have been reported by \citet{kocevski2023}: CEERS-1670 at $z = 5.2$, and CEERS-3210 at $z = 5.6$,
the latter being a heavily obscured AGN caught in a transition phase between a dust-obscured starburst 
and an unobscured quasar, similar to the transitioning quasar GNz7q at $z \sim 7.19$ reported by \citet{fujimoto2022}.
\citet{ubler2023} identified an AGN in the compact galaxy GS-3073 at $z = 5.6$, estimating a BH mass of
$\sim 10^8 M_\odot$. \citet{larson2023} reported a broad H$\beta$ emission line in CEERS-1019, a galaxy at $z = 8.7$, suggesting a BH mass of  $M_{\rm BH} \sim 10^7 M_\odot$, although their detection has a lower significance ($2.3 \sigma$).
At even higher redshift, \citet{maiolino2023a} argued that
the $z = 10.6$ galaxy GNz-11 \citep{tacchella2023,bunker2023} hosts a nuclear BH with mass 
$M_{\rm BH} \sim 2 \times 10^6 M_\odot$ accreting at super-Eddington rates.
The new extensive analysis of JWST/NIRSpec observations of GNz11 presented in \citet{maiolino2023a}
suggests extremely high gas densities, typical of the BLR of AGNs, together with other AGN-like features (ionization lines typical of AGNs, ionization cones, and a Lyman-$\alpha$ halo similar to that observed in quasars, see \citealt{maiolino2023b, scholtz2023}), while Wolf Rayet (WR) 
stars alone \citep[][]{senchyna2023} can not account for many of the observed spectral properties. 

In addition to GNz11, Maiolino et al. (in prep) have identified 10 AGNs at $4 < z < 8$ in three tiers of the JADES survey through the detection of broad Balmer emission lines, with also possible candidate merging BH systems, as traced by two broad H$\alpha$ components.
The inferred BH masses in these AGN sample are in the range between $5 \times 10^7 M_\odot$ and down to $3 \times 10^5 M_\odot$, 
probing the regime expected for heavy BH seeds formed by the collapse of Super Massive Stars (SMSs) \citep{ferrara2014}.

A statistical census of broad H$\alpha$ emission line systems among 185 galaxies at $3.8 \leq z \leq 8.9$ 
confirmed with NIRSpec has led to the discovery of 10 type-1 AGNs (corresponding to $\sim 5 \%$ of the sample) 
with masses $M_{\rm BH} \sim 10^6 - 10^7 M_\odot$ at $z = 4 - 6.9$ \citep{harikane2023}, 
two of which are the systems reported by \citet{kocevski2023}.  It is important to note that broad emission 
lines can also be powered by mergers or outflows, but in these cases broad lines are seen both in 
permitted and forbidden lines, unlike these JWST sources, whose spectra show the typical narrow–forbidden/broad–
permitted duality of AGNs. 

Additional high redshift faint AGN candidates have been identified by means of high-resolution imaging and spectral energy distribution (SED) modeling \citep{endsley2023, furtak2022, onoue2023, labbe2023}, including a system reported by \citet{bogdan2023}; this is a luminous X-ray source in the gravitationally lensed galaxy UHZ1, which has a solid photometric redshift of $z \simeq 10.3$ \citep{castellano2022}. They interpret the source as a $10^7 - 10^8 M_\odot$ accreting BH, with a bolometric luminosity of $L_{\rm bol} = [2 - 8] \times 10^{45}$ \, erg/s. Interestingly, the estimated stellar mass of the galaxy is $M_{\rm star} = 0.4^{+1.9}_{-0.2} \times 10^8 M_\odot$ \citep{castellano2022}\footnote{We note here that the stellar mass of the system has been inferred using galaxy SED templates, i.e. with no AGN contribution. However, the contribution of the AGN to the observed photometry should be subdominant, given the large gas column density ($N_{\rm H} \simeq 10^{24} \, {\rm cm}^{-2}$) inferred by the analysis of the X-ray data \citep{bogdan2023}.}, suggesting that system maybe experiencing the Outsize Black Hole Galaxy (OBG) stage predicted by theoretical models for high redshift galaxies which are the birth places of heavy BH seeds \citep{agarwal2013, natarajan2017unveiling, valiante2016}. 

These candidate AGNs, are often selected with red colors at long wavelengths, interpreted as reddened AGN emission, and a bluer component a short wavelength, possibly associated with scattered light from the AGN or with faint unobscured star formation in the host \citep{furtak2022, labbe2023}. However, the possibility of compact dusty star formation can not be ruled out, and it would be essential to unambiguously identify the presence of a faint AGN through broad wing Balmer emission-lines. 

Recently, \citet{matthee2023} have conducted a systematic search for broad line H$\alpha$ emitters over $\simeq 230 \, {\rm arcmin}^2$ covered by the EIGER \citep{kashino2023} and FRESCO \citep{oesch2023} JWST programs, corresponding to a volume of $6 \times 10^5 {\rm Mpc}^3$ over $4 \leq z \leq 6$. They reported the identification of 20 systems at $z = 4.2 - 5.5$ that have broad H$\alpha$ components with line widths from $1200 - 3700$ km/s, that they interpreted as being powered by accretion onto SMBHs with 
masses $\sim 10^7 - 10^8 M_\odot$, measuring a number densities of $\simeq 10^{-5} {\rm cMpc}^{-3}$ in the UV luminosity range 
$ -21 \leq M_{\rm UV} \leq -18$.

Given the high rate of faint AGNs discovered with JWST, two questions naturally arise:  \\
Are we surprised to find these {\it hidden little monsters}\footnote{We use here a definition taken from the title of the paper by \citet{kocevski2023}, which reads {\it Hidden Little Monsters: Spectroscopic Identification of Low-Mass, Broad-Line AGN at $z>5$
  with CEERS.}} in the hearth of galaxies at $z \sim 5  - 11$?
And can we use their estimated properties to constrain the nature and growth of the first BH seeds?

\section{Great expectations and amazing confirmations}
\label{sec:question1}

The recent discovery of more than 40 new faint AGNs (with luminosities $< 10^{46}$ erg/s) at $z>4$ in ongoing surveys with JWST, including two tentative systems at $z = 8.7$ and $10.3$, and the currently most distant AGN at $z = 10.6$, is an amazing confirmation of the rich BH landscape at cosmic dawn predicted by theoretical models.

The existence of hundreds of quasars at $6 \leq z \leq 7.64$ (see \citealt{fan2022} for a recent review, and \citealt{banados2018, yang2020} and \citealt{wang2021} for the three systems that mark the current quasars redshift frontier at $z = 7.54, 7.52$ and 7.64, respectively) show that $10^9 -10^{10} M_\odot$ super massive black holes (SMBHs) are able to form in $\sim 700$ Myr, by efficiently growing mass onto smaller black hole {\it seeds} (see \citealt{volonteri2021} and \citealt{inayoshi2020} for recent reviews). 

The nature of BH seeds is still largely unconstrained. Astrophysical BH seeds have been proposed to form in different flavours: {\it light} BH seeds with masses up to a few $\sim 100 \, M_\odot$ may be the remnants of massive Population III (Pop III) stars; {\it heavy} BH seeds with masses of $\sim 10^4 - 10^5 M_\odot$ may form by the direct collapse of a Super Massive Star (SMS); {\it medium weight} BH seeds, with masses $\sim 10^3 M_\odot$ may result from runaway stellar or black hole collisions in dense stellar clusters. The physical conditions enabling the formation of BH seeds are tightly linked to gas cooling, heating, and accretion in metal-poor environments, and their occurrence goes hand-in-hand with the formation of early cosmic structures. 

Theoretical models which attempt to incorporate a self-consistent BH seeding prescription find that the genealogy of SMBHs with masses $> 10^9 M_\odot$ powering quasars at $z \sim 6$ is characterized by a rich variety of BH seed progenitors \citep[see e.g.][]{volonteri2009, valiante2016, ricarte2018b, spinoso2022}. Only a small fraction ($< 10 - 20$\%) of BH seeds efficiently grow and contribute
to the mass of the most massive SMBHs at $z > 6$ \citep{sassano2021}, while the vast majority experience a more quiet evolution 
and populate the low-end of the mass and luminosity function, which potentially encode information on their growth mode \citep{trinca2022}.

\subsection{Expected number of JWST detectable AGNs and challenges to their detection}

Within a scenario where SMBHs at $z \sim 6$ originate from the Eddington-limited growth of light and heavy BH seeds, \citet{valiante2018observability} find that efficiently growing heavy 
BH seed progenitors are within the sensitivity limit of NIRCam\footnote{This estimate was based on the photometric performance requirements of the instrument, namely the point source faintest fluxes that can be obtained at a signal-to-noise ratio SNR = 10 in a 10 ks integration \citep{valiante2018observability}.} out to $z \sim 16$. The number of detectable seeds depends on the survey area and on the degree of obscuration. \citet{trinca2023bh} estimate that, in a survey like CEERS, between $48$ to $175$  AGNs with BH masses in the range $M_{\rm BH} = [10^6 - 10^8] M_\odot$ would be observable at $5 \leq z < 7$, and between $8$ to $21$ at $7 \leq z \leq 10$. JADES-Deep is expected to find between $12$ to $63$ AGNs with with BH masses in the range $M_{\rm BH} = [10^4 - 10^6] M_\odot$ at $7 \leq z \leq 10$, and between $5$ to $32$ at $z \gtrsim 10$. 
These numbers are predicted to be even larger in models accounting for
phases of super-Eddington accretion \citep{madau2014a, lupi2014, volonteri2015, pezzulli2016}, 
although in this regime BH growth is very intermittent, as suggested by high-resolution hydrodynamical simulations \citep{regan2019, massonneau2023, sassano2023}. 

Although these numbers are still significantly larger than the confirmed AGNs (\citealt{harikane2023, larson2023, maiolino2023a, matthee2023}), spectroscopy in these fields is still incomplete and does not yet match the depth of imaging. In addition, the theoretical predictions reported above have been estimated only on the basis of the flux limits and area coverage of ongoing JWST surveys, without estimating the contamination of the host galaxy and how this hampers the AGN identification. A common feature of faint AGNs discovered by JWST is that their identification relied on the detection of broad hydrogen (and in some cases helium) permitted lines, while their colors are hardly distinguishable from a metal-poor star forming galaxies, and often dominated by the stellar continuum of their host galaxies. Prior to the spectroscopic confirmation by \citet{kocevski2023}, CEERS-1670 was suggested to be either a strong broad-line emitter or a super-Eddington accreting BH by \citet{onoue2023} (the source was named CEERS-AGN-z5-1 in this study), on the basis of spectral fitting of deep NIRCam photometry (complemented by additional CFHT, HST and Spitzer photometric data) and its point-source morphology; however, the possibility of an extremely young galaxy with moderate dust attenuation was not completely ruled out. In addition, \citet{harikane2023} report that high-resolution HST and JWST/NIRcam images of 7 out of 10 faint AGNs identified as broad H$\alpha$ emission line systems show extended morphologies, indicating
a significant contribution to the total light from their host galaxies. Hence, commonly adopted color-color selection techniques that have been developed 
\citep{barrow2018, natarajan2017, volonteri2017, valiante2018observability, trussler2022, goulding2022} would not be able to identify these systems, unless the BH is overmassive and the AGN provides a dominant contribution to the total emission of the galaxy \citep{volonteri2023}. In addition, \citet{kocevski2023, ubler2023, larson2023}, and \citet{harikane2023} show that these JWST discovered AGNs would not be classified as such using standard spectral diagnostic techniques based on narrow line ratios that are commonly adopted to discriminate galaxies dominated by the AGN emission or by stellar emission, such as the BPT diagram \citep{baldwin1981} or the OHNO diagram \citep{backhaus2022}. 
In fact, in the BPT diagram they occupy the region with high [OIII]/H$\beta$ and low [NII]/H$\alpha$,
similar to other high-$z$ galaxies without broad line emission \citep{nakajima2023, sanders2023a} and 
within the $z \sim 0$ classical star forming regions. Similarly, in the OHNO diagram, the systems lie in the AGN region of the diagram but overlap with other high-$z$ star forming galaxies \citep{trump2023}. The observed line-ratios of all these systems are consistent with what predicted by photo-ionization models with low metallicity and high ionization parameters, and alternative diagnostic diagrams have been proposed to separate stellar emission from AGN emission in these early galaxies, such as the EW(HeII$\lambda$4686) vs HeII$\lambda$4686/H$\beta$ diagram \citep{nakajima2022}, where the source GS-3073 is compatible with an AGN and clearly separated from star forming galaxies \citep{ubler2023}.  

What are the physical properties of JWST detectable BH seeds, and can we identify - among the simulated samples - systems that could represent synthetic counterparts of the currently most distant AGN, GNz11 at $z = 10.6$ \citep{maiolino2023a}, and of CEERS-1019 at $z = 8.7$, a system that may potentially host an AGN \citep{larson2023}?

\subsection{Expected properties of JWST detectable AGNs}
\label{sec:properties}

We base our investigation on the Cosmic Archaeology Tool (CAT) model presented in \citet{trinca2022} (hereafter T22). CAT is a semi-analytical model that 
allows us to follow the co-evolution between nuclear BHs and their host galaxies all the way from the formation of the first stars and their BH remnants in the star forming molecular-cooling halos at $z \gtrsim 20 - 30$, down to the assembly of systems hosting bright quasars and less luminous AGNs at $4 \leq z \lesssim 7.5$.
The model runs on a sample of dark matter (DM) halo merger histories at $4 \le z \le 30$ generated through the galaxy formation model GALFORM \citep{cole2000, parkinson2008} based on the Extended Press Schechter theory (EPS)\footnote{We assume a Lambda cold dark matter ($\Lambda$CDM) cosmological model with the following parameters: $\Omega_\Lambda = 0.685$, $\Omega_{\rm M} = 0.315$, $h = 0.674$, $\Omega_{\rm b} = 0.05$ \citep{planck2018}.}. We adopt a redshift dependent mass resolution that corresponds to a DM halo virial temperature of 1200 K, and we simulate the formation histories of $[10^9-10^{14}]\, M_\odot$ DM halos at $z = 4$ by sampling their mass interval into 11 logarithmically spaced bins, and by generating - for each bin - 10 independent merger trees. 

In each halo, we follow gas infall, cooling, and star formation following the same implementation described in T22. Pop III stars form in metal poor star forming regions with metallicity $Z < Z_{\rm cr} = 10^{-3.8} Z_\odot$, with a mass that is stochastically sampled from a Larson initial mass function (IMF) with a characteristic mass of $m_{\rm ch} = 20 M_\odot$ in the mass range $10 M_\odot \leq m_* \leq 300 M_\odot$. When $Z \geq Z_{\rm cr}$, Pop II/I stars form according to a Larson  IMF with a characteristic mass of  $m_{\rm ch} = 0.35 M_\odot$ in the mass range $0.1 M_\odot \leq m_* \leq 100 M_\odot$. Stars are evolved according to their characteristic evolutionary timescales, and we account for their radiative, chemical, and mechanical feedback by computing their metallicity- and age-dependent radiative output (in the hydrogen ionizing and Lyman-Werner bands),  their mass- and metallicity-dependent metal/dust yields, and the energy deposited by core-collapse and pair-instability supernova explosions. This allows us to follow the gas, metal, and dust content in each galaxy, and to estimate the effects of photo-ionization and photo-dissociation feedback, which regulate the star formation efficiency in molecular-cooling halos.

The adopted BH seeding prescriptions are the same as in T22: light BH seeds form as remnants of Pop III stars, and their masses ranges from 10s to 100s $M_\odot$, depending on the Pop III star formation efficiency. Following each Pop III star formation episode, the most massive among BH remnants is assumed to migrate to the center of the host halo, becoming its nuclear light BH seed.
Heavy BH seeds are assumed to form with a constant mass of $10^5 M_\odot$ from the collapse of SMSs in atomic-cooling halos ( $T_{\rm vir} \ge 10^4$ K), when metal and dust cooling is still inefficient ($Z \leq Z_{\rm cr}$), and when molecular cooling is suppressed by a strong illuminating LW flux\footnote{The latter condition is usually expressed as
$J_{\rm LW} \geq J_{\rm cr} = 300$, where $J_{\rm LW}$ is the cumulative flux into the Lyman-Werner energy band in units of $10^{-21}$ erg s$^{-1}$ cm$^{-2}$ Hz$^{-1}$ sr$^{-1}$ (see T22).}. Hence, in CAT heavy and light seeds co-exist and either of the two types of seed forms when the physical conditions allow. This means that the BH population predicted at each given redshift descends from both light or heavy BH seeds, with the exact match at late times determined by the ability or inability of the seeds to grow, as described in the following.  \\

Once formed, BH seeds can grow by gas accretion and mergers. In particular, following galaxy major mergers (defined as those with interacting halos mass ratio greater than 1/10), the nuclear BHs are assumed to merge with their host galaxies. In doing so we do not account for the (admittedly poorly constrained, see \citealt{bortolas2020}) dynamical friction timescale, as well as for the subsequent binary hardening phase, and kick velocity of the newly formed BHs. All these processes may have important effects on the BH occupation fraction and on the estimated BH binary merger rates (see the discussion in \citealt{volonteri2021}). While an implementation of physically motivated prescriptions to account for these processes in CAT is currently underway (Trinca et al. in prep), we do not expect these processes to significantly affect our predictions, given that nuclear BHs grow predominantly by gas accretion rather than by mergers. 

Here we consider CAT predictions for two alternative BH accretion models: in the first model, the accretion onto light and heavy BH seeds is Eddington-limited (EL, the model referred to as the reference model in T22); in the second model, light and heavy BH seeds can accrete at super-Eddington rates in short phases triggered by galaxy major mergers (SE, the model referred to as merger driven model in T22). The maximum duration of the SE phases is 1/10 of the host halo dynamical timescale ($\Delta t_{\rm burst} \simeq 2 - 7$ Myr at $z \simeq 10 - 20$, see section 2.4.2 in T22), but SE accretion is generally sustained for $\lesssim 2 - 5$ Myr, consistent with predictions based on high-resolution hydrodynamical simulations \citep{regan2020, sassano2023, massonneau2023}. When computing the BH bolometric luminosity, in the EL model we take a fixed radiative efficiency of $\epsilon_{\rm r} = 0.1$, while in the SE model we adopt a slim disk solution with the parametric form proposed by \citet{madau2014a}, and we assume a constant BH spin of 0.572, so that when the BHs are not accreting at SE rates their $\epsilon_{\rm r} \rightarrow 0.1$. 

CAT has been used to explore observational imprints of light and heavy BH seeds and their growth mode on the redshift evolution of the BH mass and luminosity functions (see T22), their detectability within ongoing JWST surveys \citep{trinca2023bh}, the nature of UV-bright galaxies detected by JWST at $z \ge 10$ \citep{trinca2023gal}, and the expected imprints of the first stellar populations and accreting BH seeds on the global 21cm signal at cosmic dawn \citep{ventura2023}. 

\begin{figure*}
    \centering
    \includegraphics[width=16cm]{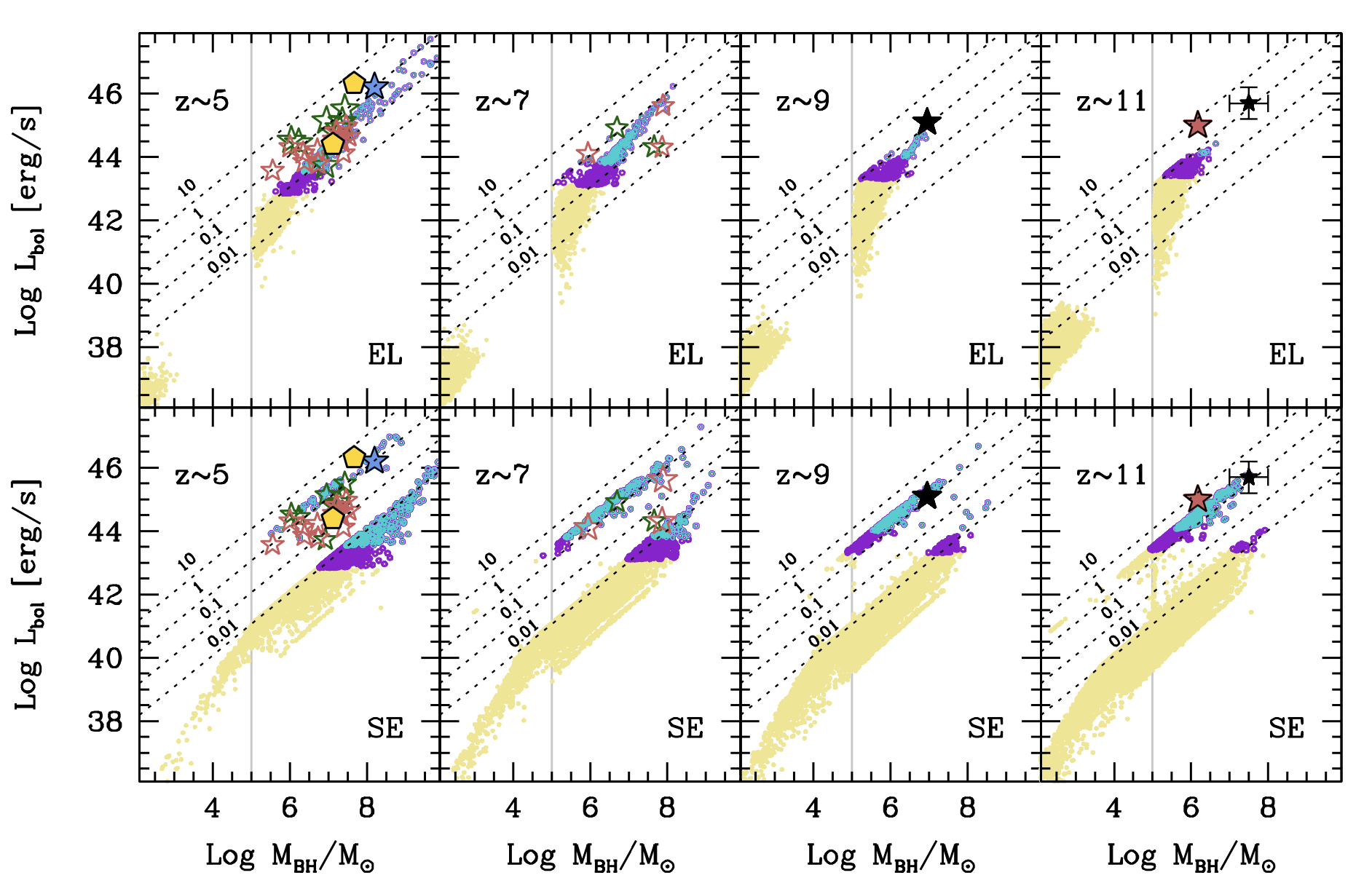}
    \caption{BH masses and bolometric luminosities predicted by CAT (T22) at $5 \lesssim z \lesssim 11$ assuming the Eddington limited model (EL, top row) and the super-Eddington model (SE, bottom row), where BHs can accrete at super-Eddington rates in short phases triggered by galaxy major mergers.  
    In each panel, light yellow points represent all the systems at the corresponding redshift, which can descend from light and/or heavy BH seeds. Systems with a luminosity above the limiting sensitivity of JADES and CEERS are circled in violet and turquoise, respectively. The diagonal dashed lines show reference values of $\lambda_{\rm Edd} = L_{\rm bol}/L_{\rm Edd} = 0.01, 0.1, 1,$ and 10, and the vertical gray lines mark the initial heavy BH seeds mass adopted in CAT ($10^5 M_\odot$). We also show the luminosities and BH masses identified in JWST surveys by \citet{kocevski2023} (yellow pentagons at $z \sim 5$), \citet{ubler2023} (blue star at $z \sim 5$), \citet{harikane2023} (green hollow stars at $z \sim 5$ and $\sim 7$); the candidate AGN in CEERS-1019 identified by \citet{larson2023} is marked with a black star at $z \sim 9$, the AGN in GNz11 reported by \citet{maiolino2023a} is shown as a filled red star at $z \sim 11$, and AGNs at lower redshift identified by Maiolino et al. (2023c) are shown by hollow red stars. In the rightmost panels we also show as a black star with errorbars the candidate AGN at $z \simeq 10.3$ reported by \citet{bogdan2023}.
    The properties of some of the observed systems at $z \sim 5$ are consistent with predictions of the EL model, but the luminosity and BH masses of the remaining systems at $z \sim 5$ and of most of the systems at $z \gtrsim 7$ are better reproduced in the SE model.}
    \label{fig:mbhlbol}
\end{figure*}

\begin{figure*}
    \centering
    \includegraphics[width=16cm]{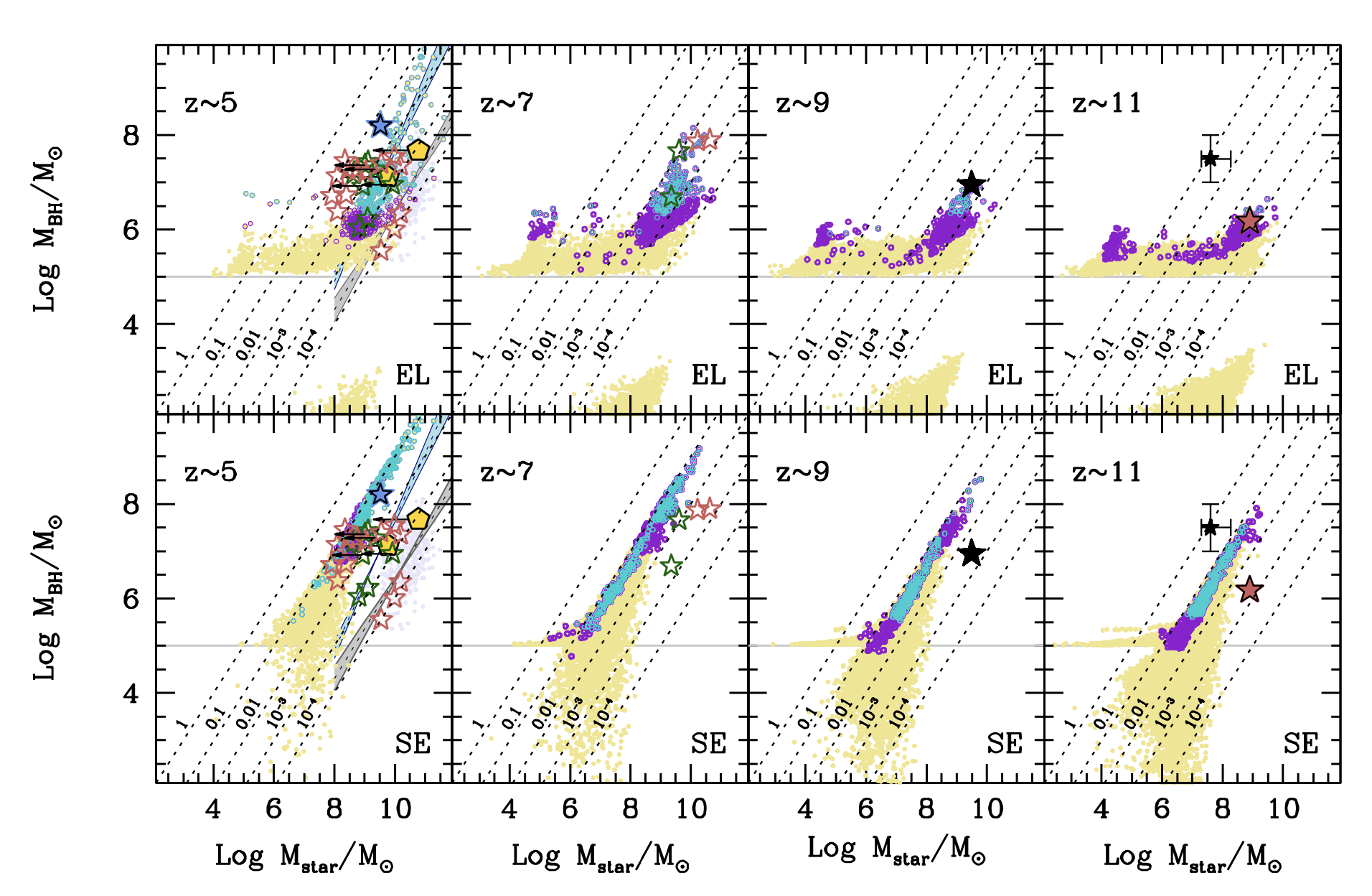}
    \caption{Similar to Fig. \ref{fig:mbhlbol}, but showing the BH mass - stellar mass relation predicted by CAT (T22) at $5 \lesssim z \lesssim 11$ for the Eddington limited model (EL, top row) and the super-Eddington model (SE, bottom row).
     The diagonal dashed lines show reference values of $M_{\rm BH}/M_{\rm star}$ from $10^{-4}$ to 1, and the horizontal gray lines mark the initial heavy BH seeds mass adopted in CAT ($10^5 M_\odot$). The gray and azure shaded areas in the $z \sim 5$ panels show, respectively, the fit of from \citet{reines2015} to their local sample of broad-line AGNs (light purple points), and the updated fit to the scaling relation from \citet{greene2020}. We also show the estimated BH and stellar masses (or upper limits) of systems identified in JWST surveys by \citet{kocevski2023} (yellow pentagons at $z \sim 5$), \citet{ubler2023} (blue star at $z \sim 5$), \citet{harikane2023} (green hollow stars at $z \sim 5$ and $\sim 7$); the candidate AGN in CEERS-1019 identified by \citet{larson2023} is marked with a black star at $z \sim 9$, the AGN in GNz11 reported by \citet{maiolino2023a} is shown as a filled red star at $z \sim 11$, and AGNs at lower redshift identified by Maiolino et al. (2023c) are shown by hollow red stars. In the rightmost panels we also show as a black star with errorbars the candidate AGN at $z \simeq 10.3$ reported by \citet{bogdan2023}. At all redshifts, most of the observable systems have $M_{\rm BH}/M_{\rm star} \lesssim 0.01$ in the EL model, and $0.01 \lesssim M_{\rm BH}/M_{\rm star} \lesssim 0.1$ in the SE model.}
    \label{fig:mbhmstar}
\end{figure*}
BH masses and luminosities of AGNs detectable by JADES and CEERS at $5 \lesssim z \lesssim 11$ are shown in Fig. \ref{fig:mbhlbol}. In each panel, the BH population grown from light and heavy BH seeds at the given redshift is shown by yellow points, while systems with luminosity brighter than the limiting magnitude of  JADES and CEERS surveys are circled in violet and turquoise, respectively. Hence, turquoise points mark systems that can be observed by both surveys, while violet points mark systems that can only be detected by JADES. The upper and lower panels show, respectively, the systems predicted by CAT in the EL and SE models. \\
At all redshifts, the systems detectable by JADES extend to lower BH masses and luminosities, as expected by the fainter limiting magnitude of the survey compared to CEERS \citep{trinca2023bh}. In both models, the bolometric luminosity increases with BH mass, but with some discontinuities: in the EL model, BHs descending from light seeds grow very inefficiently, and their masses and luminosities remain $\lesssim 10^4 M_\odot$ and ${\rm Log} (L_{\rm bol}) [\rm erg/s] \lesssim 39.5$ at all redshifts. Above this mass and luminosity, there appears to be a mass gap, as - by construction - the minimum heavy BH seed mass adopted in CAT is $M_{\rm BH} = 10^5 M_\odot$ (gray vertical lines). While this gap could be filled if a more continuous distribution of initial BH seeds masses were adopted (see e.g. \citealt{ferrara2014, sassano2023})\footnote{Medium weight BH seeds with masses $\sim 10^3 M_\odot$ are predicted to form by runaway collisions in the dense stellar clusters \citep{omukai2008, devecchi2009, devecchi2012}, and a more continuous distribution between medium weight and heavy BH seeds is found within the super-competitive accretion model recently explored by \citet{chon2020}.}, in the present version of the EL model all the systems detectable with JADES and CEERS appear to be descendants of heavy BH seeds (see T22 for a discussion on the consequences of the adopted accretion model on the evolution of the BH mass and luminosity functions). 

In the SE model, even light BH seeds grow more efficiently and the mass gap disappears\footnote{We note here that in our merger-driven SE accretion model light BH seeds grow more efficiently than predicted by other studies (see e.g. \citealt{smith2018}). We will return to this point in section \ref{sec:conclusions}.}. However, there is a luminosity gap, with two disjoint, parallel diagonal tracks, reflecting the short duration of the SE phase, which shift part of the systems on the upper track. The comparison with the diagonal dashed lines shows that systems on the upper track are characterized by Eddington ratios in the range $1 \lesssim \lambda_{\rm Edd} = L_{\rm bol}/L_{\rm Edd} \lesssim 10$, while systems on the lower track have $\lambda_{\rm Edd} \lesssim 0.01 - 0.1$, smaller than in the EL model.
The upper track becomes progressively less populated with cosmic time, due to the decreasing number of galaxy major mergers and their decreasing gas content (see also \citealt{pezzulli2016} for a similar finding). 

As expected, in the SE model the number of detectable systems is larger than in the EL model, particularly at the highest redshift, where most of the systems are detectable when accreting at SE rates. Since BH growth is faster in the SE model, CEERS and JADES are predicted to observe systems with a broader range of BH masses, at all redshifts, with JADES reaching the limit of the initial heavy BH seed mass at $z \lesssim 11$, with a few systems even crossing this limit at $z \sim 7 - 9$, hence potentially probing BHs originated from lighter BH seeds. At lower redshifts, SE accreting BHs with masses $\sim 10^5 M_\odot$ become progressively rarer, as these systems have already grown, loosing memory of their initial BH seed mass \citep{valiante2018observability}. Among the sample of AGNs detected by JWST, the BH mass and luminosity of a systems at $z \sim 5$ are broadly consistent with predictions of the EL model, but most of the sources at $z \geq 7$ are better reproduced by the SE model.

The distribution of simulated systems in the BH mass - stellar mass plane for the EL and the SE models is shown in Fig. \ref{fig:mbhmstar}, using the same color-coding adopted in Fig.\ref{fig:mbhlbol}. In the $z \sim 5$ panels, we also show the sample of local broad-line AGNs (light purple data points) from \citet{reines2015}, and their best fit relation (gray shaded region), together with the updated fit to the larger sample from \citet{greene2020} (azure shaded region). We stress that - while pre-JWST observations based on dynamical mass measurements of $z \simeq 6$ quasars suggested a strong evolution of the relation with redshift (see \citealt{fan2022} for a recent review) - recent NIRCam/NIRSpec observations of two quasar at $z \sim 6 - 7$ show that their location in the black hole mass - stellar mass plane is consistent with the distribution at low
redshift \citep{ding2022}, suggesting that the most massive BHs inhabit the most massive galaxies\footnote{We note here that the inferred relation at $z \simeq 6$ is strongly affected by potential selection biases and by using gas tracers \citep{volonteri2011, lupi2019}, and that massive quasar hosts were indeed expected by evolutionary scenarios that attempt to reproduce their inferred large dust masses \citep{valiante2014}.}. 

Figure \ref{fig:mbhmstar} shows that the BH mass gap is apparent in the EL model, where light BH seeds descendants with masses $< 10^4 M_\odot$ are undermassive with respect to their host galaxies. Even for the more efficiently growing heavy BH seeds, their masses remain $\lesssim 10^6 M_\odot$ until the host galaxy has grown to a mass of at least $M_{\rm star} > 10^8 M_\odot$, above which the growth of the BHs adjust to $M_{\rm BH}/M_{\rm star} \lesssim 0.001$, with a large scatter (see also \citealt{habouzit2017} and \citealt{alcazar2017}). Interestingly, a small fraction of systems observable with JADES at $z \gtrsim 7$ is made by heavy BH seeds with $\lesssim 10^6 M_\odot$ which are over-massive with respect to their host galaxies, as expected if these systems keep memory of their 
peculiar birth conditions \citep{agarwal2013, natarajan2017, visbal2018, valiante2018observability}.

In the SE model, instead, the vast majority of detectable BHs grow at a pace almost similar to that of their hosts, with $M_{\rm BH}/M_{\rm star} \simeq 0.01 - 0.1$; while this result might reflect the relatively simplified treatment of the SE accretion in CAT (see the discussion in T22 and in section \ref{sec:conclusions}), it certainly shows how the assumed BH growth translates into a different evolution of the BH mass - stellar mass relation, even among models which rely on the same BH seeding conditions \citep{volonteri2012}. 
 
At all redshifts, most of the observable systems have $0.001 \lesssim M_{\rm BH}/M_{\rm star} \lesssim 0.01$ in the EL model, and $0.01 \lesssim M_{\rm BH}/M_{\rm star} \lesssim 0.1$ in the SE model.  
The comparison with JWST detected AGNs shows that their luminosities are better reproduced by the SE model, while their host galaxy masses are better 
matched by the EL model, particularly at $z \gtrsim 7$. We also note that if the estimated photometric redshift, BH mass and host galaxy stellar mass of UHZ1 at $z = 10.3$ were to be confirmed by future observations \citep{castellano2022, bogdan2023}, this system - shown by the small black star with errorbars in the rightmost panel of the figure - would be more overmassive (in terms of BH to stellar mass ratio) than predicted by both models.

We nevertherless identify the subsample of detectable sources with BH masses that match the measured BH masses estimated for GNz11 \citep{maiolino2023a} and CEERS-1019 \citep{larson2023}, and whose stellar mass (for sources in the EL model) or BH luminosity (for sources in the SE model) is consistent with the observed values. For these subsample of systems, which provide our best synthetic analogues for GNz-11 and CEERS-1019, we compute below the emission spectra and analyse their evolutionary tracks. 

\subsection{Emission spectra of the first accreting BHs and their hosts}

The best synthetic candidates for GNz11 at $z = 10.6$ and CEERS-1019 at $z = 8.68$ are extracted from the source catalogs predicted by CAT in both the EL and SE models. We first select systems with a BH mass comparable to that estimated for the two observed sources \citep{maiolino2023a, larson2023}. 
Then, we also match two or more additional properties such as the estimated AGN bolometric luminosity, Eddington ratio ($\lambda_{\rm Edd}$), host galaxy stellar mass and metallicity \citep[][]{larson2023,  tacchella2023}. 

For CEERS-1019 we also explore the possibility that the system does not host an accreting nuclear BH, since the identification of the BLR through the broad H$\beta$ emission line has a low signal-to-noise ratio ($2.3 \sigma$). In this case, we select our synthetic candidates to have stellar masses and star formation rates within the observed range, with no additional condition on the BH mass and luminosity.

For each of these systems, we compute an incident continuum spectrum from the accreting BH and from star formation in the host galaxy, based on the properties (redshift, BH mass and bolometric luminosity, stellar mass, age and metallicity) predicted by CAT. We then use these spectra as input to Cloudy photoionization models
(we adopt version 22.00 of Cloudy \citealt{ferland2017}) to compute the transmitted 
fluxes, adopting the interstellar medium (ISM) properties predicted by CAT for each system (gas metallicity, dust-to-gas, and dust-to-metals mass ratios), similarly to what done in our previous study \citep{valiante2018observability}.

A systematic investigation of the Spectral Energy Distributions (SEDs) for all the systems predicted by CAT to be detectable by JWST (and their position in spectral diagnostic diagrams) will be presented in Valiante et al. in prep. Here we show the SEDs for candidates which minimize the difference with the photometric data of CEERS-1019 and GNz11, according to our Cloudy models. 

\begin{figure}
    \centering
    \includegraphics[width=8.5cm]{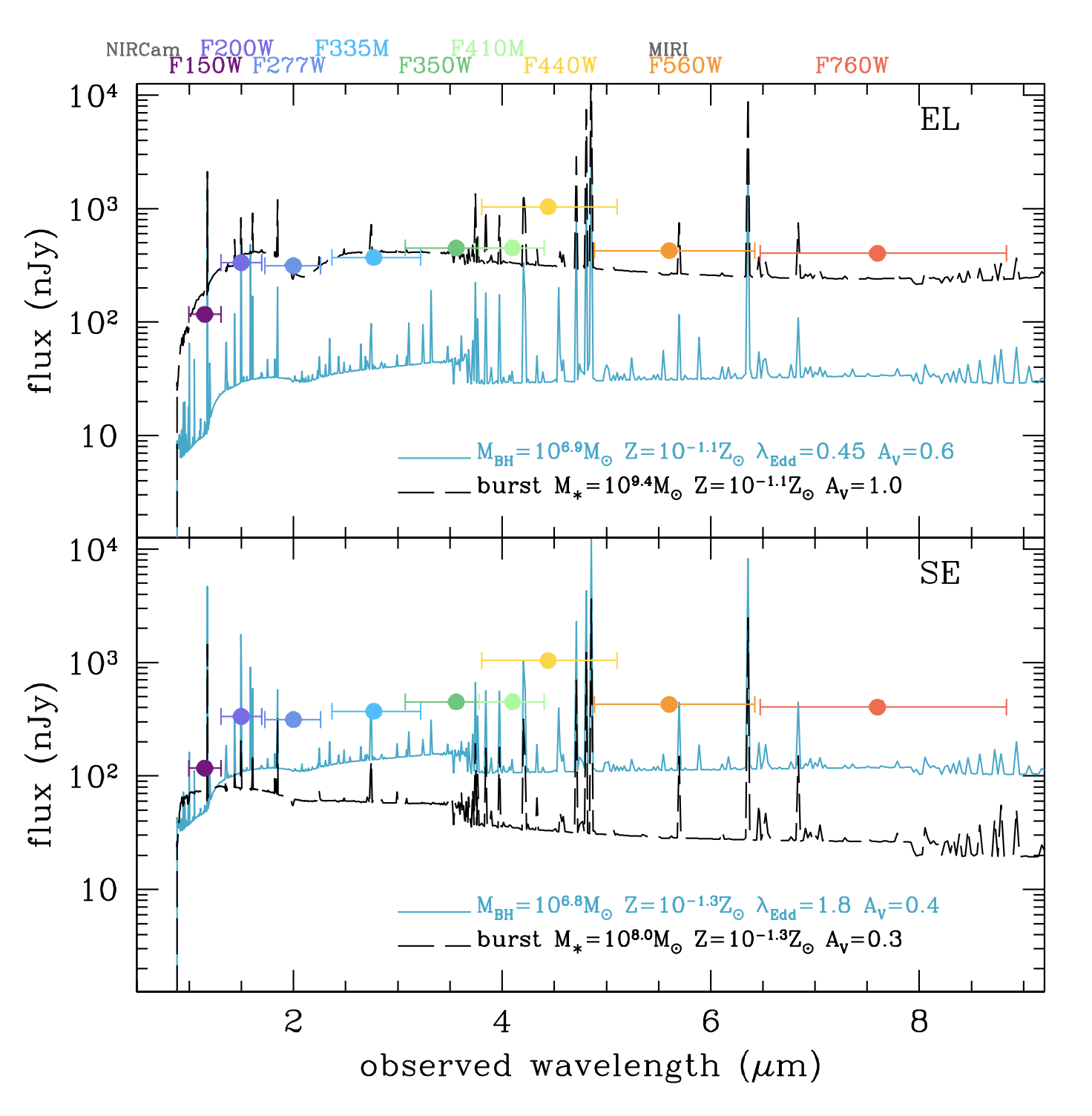}
    \caption{Flux density predicted with Cloudy for CEERS-1019 candidates selected from the EL model (upper panel) and from the SE (lower panel) model catalogs at $z \sim 9$ predicted by CAT (see section \ref{sec:properties}). The solid azure and dashed black lines in both panels show the SED of the accreting BH and stellar component, respectively, assuming that all the stars are formed in a single burst and a stellar age of 2 Myr. 
    NIRcam and MIRI photometric data points from \citet{larson2023} are shown in different colors, with bars indicating the filter band width.  It is important to emphasize that while the observed spectrum shows a broad line attributed to the BLR of the BH, Cloudy does not calculate the line broadening caused by the BH potential (see text).}
    \label{fig:ceersSED}
\end{figure}
\begin{figure}
    \centering
    \includegraphics[width=8.5cm]{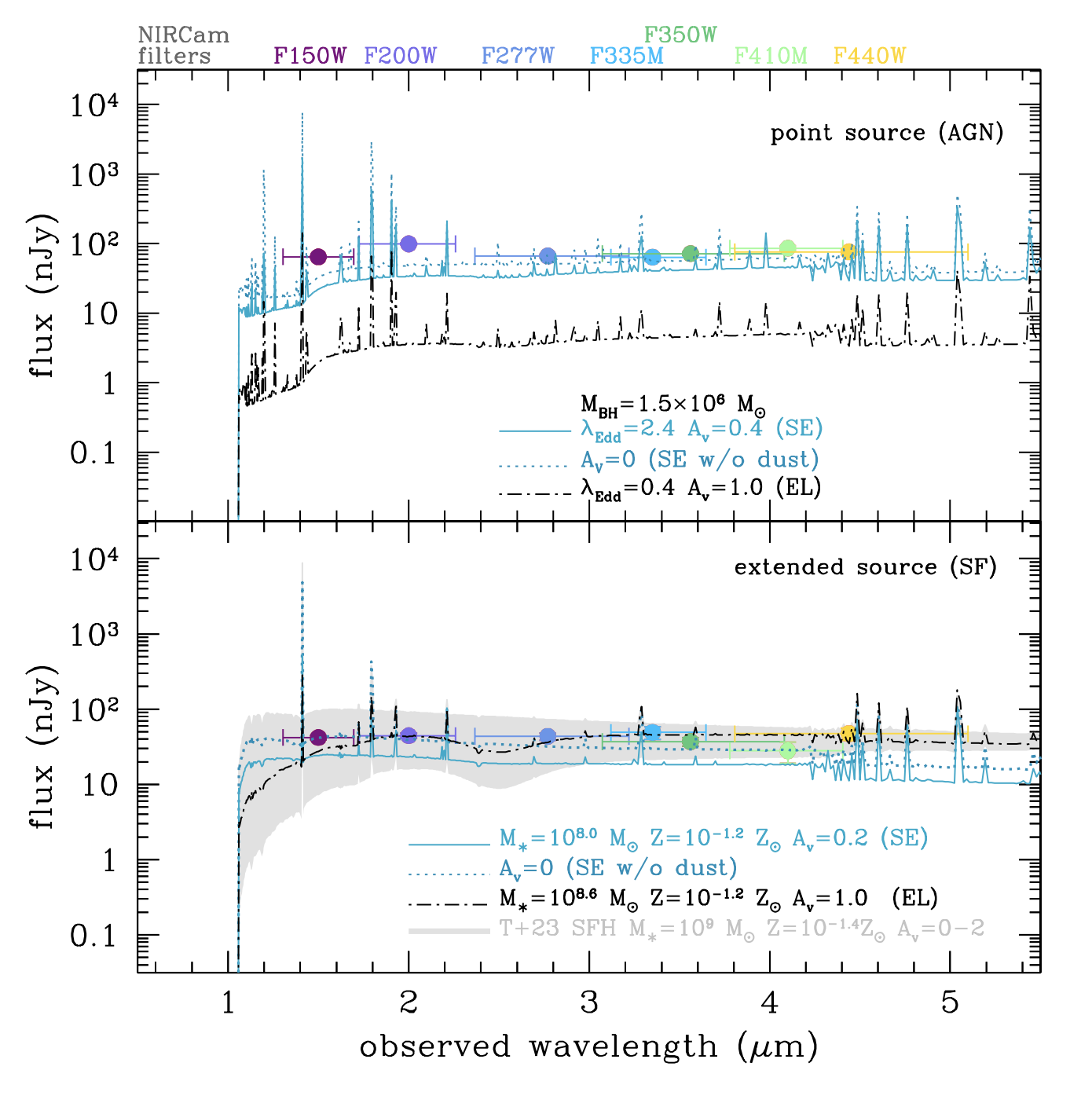}
    \caption{Flux density predicted with Cloudy for GNz11 candidates selected from the EL and SE model catalogs at $z \sim 11$ predicted by CAT (see section \ref{sec:properties}). In the upper panel we show the emission from accreting BHs and compare these to the photometric data points reported by \citet{tacchella2023} for the point source of GNz11, which is attributed to an accreting BH by \citet{maiolino2023a}. The spectra are shown for a super-Eddington ($\rm \lambda_{Edd}=2.4$) accreting BH candidate with (azure solid line, $A_{\rm V}=0.4$) and without (azure dotted line,$A_{\rm V}=0$) dust attenuation. The sub-Eddington ($\rm \lambda_{Edd}=0.4$) candidate found in the $z=11$ EL model catalog is represented by the black dot-dashed line. For this latter system we predict the same metallicity as for the SE model, 
    $0.06 Z_\odot$ (from CAT), but $A_{\rm V}=1.0$ (from the adopted Cloudy model). For all these systems the BH mass is $1.5\times 10^{6} \, \rm M_\odot$.  
    Using the same colour coding, in the bottom panel we show the emitted spectra of the corresponding host galaxies, whose stellar mass, metallicity and dust attenuation are reported in the legenda. For all these models, we have assumed all the stars to form in a single burst and a stellar age of 1 Myr. The gray shaded region shows the flux density predicted for a ${\rm log}(M_{\rm star}/M_\odot) = 9$ galaxy adopting the star formation history inferred by \citet{tacchella2023} for the extended component, assuming a fixed metallicity of $Z = 10^{-1.4} Z_\odot$, and a dust attenuation in the range $A_{\rm V} = 0 - 2$. The model predictions are compared to the photometric data points reported by \citet{tacchella2023} for the extended component of GNz11. It is important to emphasize that while the observed spectrum shows a broad line attributed to the BLR of the BH, Cloudy does not calculate the line broadening caused by the BH potential (see text).}     
    \label{fig:gnz11SED}
\end{figure}

\subsubsection{Construction of the SEDs}
The adopted models for the intrinsic emission spectra of accreting BHs and their hosts (used to define the incident radiation in Cloudy), and the main set-up conditions adopted to run Cloudy models are briefly described below. 

Following \citet{valiante2018observability}, the primary optical/UV emission from the BH accretion disk is described as the sum of multicolour black-body spectra, normalized to the BH bolometric luminosity. The X-ray emission from the hot corona is modelled as a power-law with an exponential cut-off, $L_\nu \propto \nu^{-\Gamma+1} {\rm e}^{-h\nu/E_{\rm c}}$, with $E_{\rm c} = 300$ keV and the photon index $\Gamma$ in the hard X-ray energy band is assumed to depend on the Eddington ration $\lambda_{\rm Edd}$ as $\Gamma_{\rm 2-10 keV} = 0.32 {\rm Log} \lambda_{\rm Edd} +2.27$. We also include the metallicity-dependent X-ray reflection component using the reflection-only solution of the PEXRAV model in the XSPEC code (see \citealt{valiante2018observability} for more details). 

The emission from the stellar component is computed with the spectral evolution code \textsc{PEGASE} v2.0\footnote{http://www2.iap.fr/users/fioc/Pegase/} \citep{fioc2019}, using as an input the ages and metallicities of the stellar populations formed in the host galaxy of the accreting nuclear BH predicted by CAT. We note here that when computing the stellar emission, we do not consider the full star formation history of the host galaxy, but we simply assume that 
all the stars formed in a burst and evolve over a times-scale of $\Delta t= 2$ (1) Myr, corresponding to the typical redshift-dependent time interval in CAT at $z=8-9$ ($10-11$). We compute their cumulative emission considering their metallicity to be the same as the metallicity of their parent star forming cloud and a stellar age equal to $\Delta t$.

In Cloudy models, we use the ISM properties (metallicity, dust-to-gas, dust-to-metal ratios) predicted by CAT to characterize the enrichment of the gas cloud.
For systems with metallicity and dust-to-metal mass ratio close to local interstellar values (that we assume to be $Z = Z_\odot = 0.01524$, and  $\xi_d = \xi_\odot = 0.36$, respectively), we use the interstellar abundances and depletion factors from \cite{gutkin2016}\footnote{The abundances computed by \cite{gutkin2016} also account for both primary and secondary nucleosynthesis production of Nitrogen and Carbon.}, except for Nitrogen, for which we adopt the depletion factor from \citet[][see \citealt{nakajima2018, nakajima2022}]{dopita2006}. 
For sources with sub-solar metallicity, the corresponding abundance of elements lighter than Zinc are computed by rescaling the solar values by the metallicity predicted by CAT.
In addition, following \citet{gutkin2016}, the depletion factors associated with different dust-to-metal mass ratios ($\xi_d<\xi_\odot$) are obtained with a linear interpolation of the depletion factors associated with $\xi_d=1$, 0.36 and 1. To account for dust physics, we rescale the default ISM grains abundance of Cloudy by the dust-to-gas mass ratio predicted by CAT. Grain size distribution and optical properties are from \cite{mathis1977} and \cite{martin1991}, already included in Cloudy. 

The clumping properties of the emission-line gas are defined by the filling factor (or clumping factor), $f$. For this study, we investigated both a uniform distribution model, $f=1$, and clumpy gas models with $f=0.05$, $0.1$ and $0.5$. For a given value of the dust-to-gas mass ratio (provided by CAT for each source), the adopted value of $f$ translates into a different value of the visual extinction $A_{\rm V}$. 

Following \citet{nakajima2022}, in all models presented here, we assumed a constant Hydrogen density $\rm log (n_H/cm^{-3})=3.0$ and we stop Cloudy photo-ionization runs when the electron temperature, $T_{\rm e}$, drops below 100 K (in stellar models), or when a neutral Hydrogen column density of $10^{21} {\rm cm}^{-2}$ is reached (in accreting BH models).

The resulting SEDs of systems that have been selected to be representative of the best synthetic candidates for CEERS-1019 at $z = 8.68$ and GNz11 at $z = 10.6$ are shown in Figs. \ref{fig:ceersSED} and \ref{fig:gnz11SED}. In particular, we show the SEDs that minimize the difference with the photometric data points reported by \citet{larson2023} for CEERS-1019 and by \citet{tacchella2023} for the point source of GNz11, which is attributed to an accreting BH by \citet{maiolino2023a}. It is important to 
emphasize that while the spectra of the observed sources show broad lines attributed to the BLR of the BH, Cloudy does not calculate the line broadening caused by the BH potential.

\subsubsection{CEERS-1019}
For CEERS-1019, we show the predicted emission spectra for the accreting BHs (azure solid lines) and host stellar populations (dashed black lines) of two systems, selected from the EL (upper panel) and SE models (lower panel). 
The two candidates have comparable BH mass and were selected among the closest CEERS-1019 synthetic counterparts (in terms of BH mass, luminosity and stellar mass) predicted by CAT at $z \sim 9$. 
 
In both panels, we report their main physical properties and we compare the emission spectra with NIRcam and MIRI photometric data points from \citet{larson2023}, that are shown in different colors, with bars indicating the filter band width. 

In the EL model, the continuum is dominated by the emission from the host galaxy, which has a total stellar mass of ${\rm log}(M_{\rm star}/M_\odot) = 9.4$. Here we have assumed that all stars formed in a single burst and a stellar age of 2 Myr, that is the typical time interval at $z\sim 8-9$ in CAT. Despite this extreme assumption, some of the emitted lines are heavily contaminated by the accreting BH, which has a mass of ${\rm log}(M_{\rm BH}/M_\odot) = 6.9$ and Eddington ratio of $\lambda_{\rm Edd}=0.45$. 

In the SE model, instead, the continuum and emitted lines are dominated by the BH emission, which has a mass of ${\rm log}(M_{\rm BH}/M_\odot) = 6.8$ and Eddington ratio of $\lambda_{\rm Edd}=1.8$. Due to the faster growth of the BH, the host galaxy has a stellar mass of only ${\rm log}(M_{\rm star}/M_\odot) = 8.0$. 

The metallicities of the two systems clearly reflect their different degree of evolution, being $\simeq 0.08 Z_\odot$ in the EL model, and a factor $\sim 1.6$ smaller in the SE model. Cloudy models with $f = 0.05$ provide $A_{\rm V} = 0.3 - 0.6$, in good agreement with the value inferred for CEERS-1019 \citep[$0.4\pm 0.2$;][]{larson2023}. The only exception is the host galaxy in the EL model (dashed line in the upper panel), for which we show the SED with attenuation $A_{\rm V}=1.0$, obtained assuming $f = 0.1$. For a filling factor $f < 0.1$ ($A_{\rm V} < 1.0$) the continuum emission of this stellar component exceeds NIRCam observations at $\lambda < 3.5 \mu$m.

We find that the SE model, where the flux density is dominated by the accreting BH, provides a poor fit to the observed photometry of CEERS-1019. This conclusion generally applies to SE candidates with comparable BH mass and luminosity, and supports the interpretation provided by \citet{larson2023}, i.e. that - if CEERS-1019 hosts an AGN - the emission from the host galaxy comprises between $\sim 80 - 99.5\%$ of the UV/optical flux, depending on the adopted AGN template.

We also explore the possibility that CEERS-1019 does not host an AGN, i.e. that the 2.3 $\sigma$ detection of the H$\beta$ broad line component is not confirmed by future observations. In this case, we select from CAT simulated galaxies at $z \simeq 9$ systems with stellar masses and star formation rates consistent with the values inferred by \citet{larson2023}, without imposing any additional condition on the nuclear BH mass and luminosity. In the SE model, all the selected galaxies host nuclear BHs with masses ${\rm log}(M_{\rm BH}/M_\odot) = 7.4 - 8.5$, and $\lambda_{\rm Edd} = 0.009 - 0.03$, i.e. more massive and less active than inferred by \citet{larson2023}. In the EL model, 
we find that $\simeq 10\%$ of the selected galaxies host nuclear BHs that have properties similar to the system shown in the top-panel of Fig. \ref{fig:ceersSED}, while the rest of the sample host smaller (${\rm log}(M_{\rm BH}/M_\odot) = 5.25 - 6.5$), and less luminous BHs ($\lambda_{\rm Edd} = 0.03 - 0.3$). 

\subsubsection{GNz11}
For GNz11, Fig. \ref{fig:gnz11SED} compares the predicted emission spectra powered by accreting BHs (upper panel) and their host galaxies (lower panel) with the photometric data reported by \citet{tacchella2023} for the central point source (which \citealt{maiolino2023a} attribute to the AGN emission) and the extended component, respectively. 

In the upper panel, we show the flux density predicted for two accreting BHs with mass $M_{\rm BH}\sim 1.5 \times 10^6 \, \rm M_\odot$, selected from the SE model ($\lambda_{\rm Edd} = 2.4$, azure lines) and from the EL model ($\lambda_{\rm Edd} = 0.4$, black dashed-dotted line). Using the same color-coding, in the bottom panel we show the flux density predicted for their host galaxies. The two accreting BHs are hosted by galaxies with different stellar mass (${\rm log}(M_{\rm star}/M_\odot) = 8.6$ in the EL model, and ${\rm log}(M_{\rm star}/M_\odot) = 8$ in the SE model), but with the same ISM metallicity ($Z = 0.06 Z_\odot$). 

To best match the photometry of the point-source and extended component, we performed different Cloudy runs assuming different filling factors, and found that the best match is provided by $f = 0.05$ for the SE model, corresponding to $A_{\rm V} = 0.4$ (azure solid line), and $f = 0.5$ for the EL model, corresponding to $A_{\rm V} = 1$ (the latter choice provides the best fit to the photometry of the extended component). For the SE model, we also show a model with no dust obscuration ($A_{\rm V} = 0$, azure dotted line). We find that the SED of the SE model candidate is in good agreement with the observed photometry of the point source, especially if dust extinction is neglected, while the EL model candidate is too faint.  By assuming $f=0.05$, we can reduce the visual extinction to $A_{\rm V} = 0.3$ \citep[more consistent with that inferred by][]{tacchella2023}, and the emission from the EL accreting BH increases (by a factor $\approx 3$), but it is still too small to reproduce the observations. With the same assumption, however, the emission of the stellar component is too bright to match the photometry of the extended component.

We recall that the stellar emission is computed assuming that the stars are formed in a single burst and, in this case, have an age of 1 Myr (approximately the CAT time interval at $z\sim 10-11$). To appreciate the impact of considering the contribution of older stellar populations, we also show the flux density predicted for a galaxy with total stellar mass of ${\rm log}(M_{\rm star}/M_\odot) = 9$ adopting the star formation history inferred by the SED fitting of the extended component of GNz11 by \citet{tacchella2023}, with the gray shaded region corresponding to models with different degrees of dust obscuration, in the range $A_{\rm V} = 0 - 2$. This shows that a better modeling of the star formation history of the host galaxy is required to properly account for the observations, and that the galaxies hosting the SE accreting BHs may be too faint to account for the observed flux density of the extended component, particularly at 2 - 4 $\mu$m.  

Our analysis suggests that the emission of the central point source in GNz11 can be reproduced by SE accreting BHs, with mass and luminosity comparable to the values inferred by \citet{maiolino2023a}; however, these systems are hosted in galaxies which are too faint to match the observed emission of the extended component.

\begin{figure*}
    \centering
    \includegraphics[width=16cm]{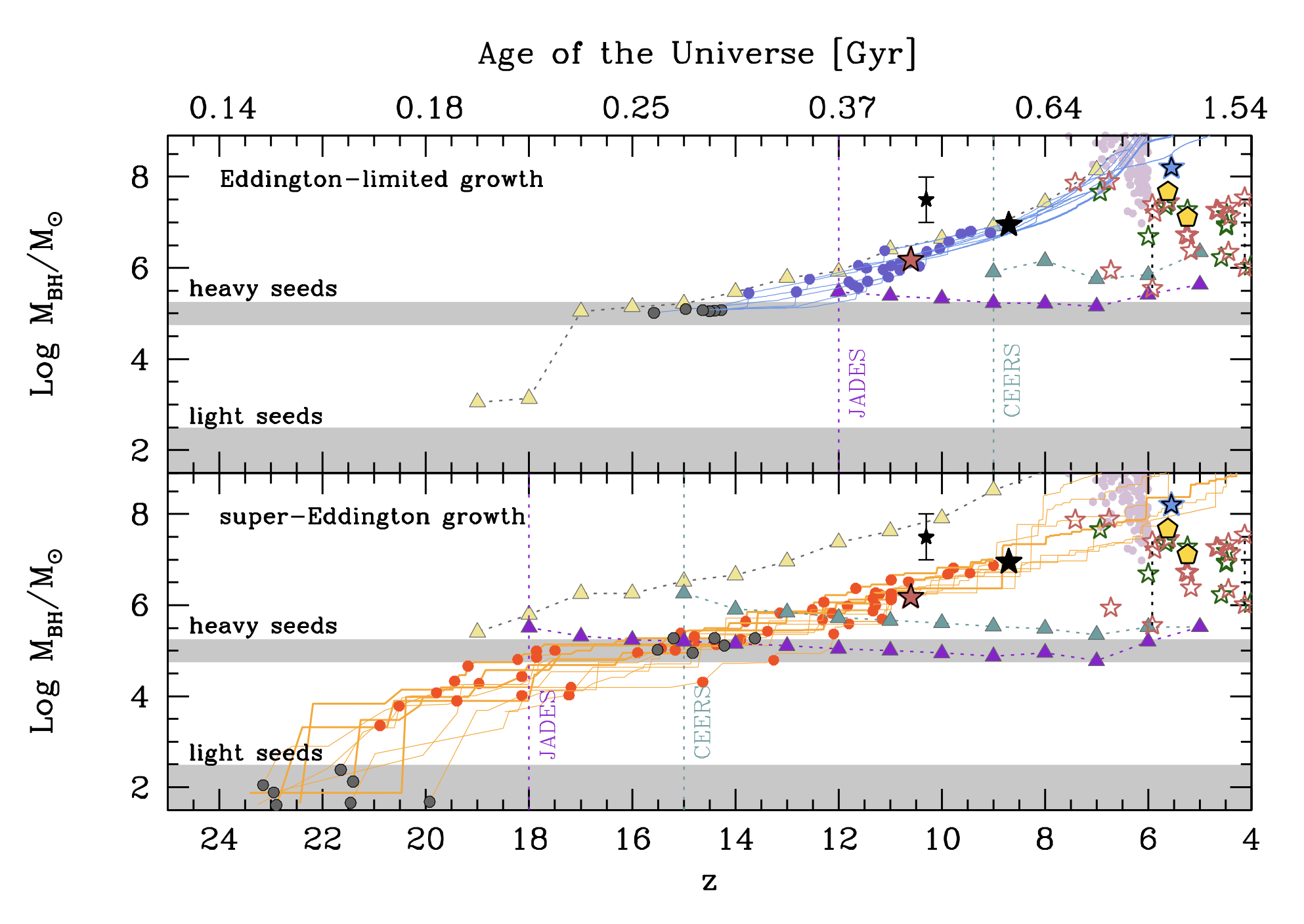}
    \caption{BHs evolutionary tracks predicted by CAT for the Eddington limited model (EL, top 
     panel) and the super-Eddington model (SE, bottom panel). In top (bottom) panel, the tracks of the best candidates for CEERS-1019 and GNz11 are shown with blue (orange) solid lines, with darkblue (red) data points indicating the occurrence of BH mergers, and the gray data points indicating the BH mass at seed formation. The horizontal gray shaded regions mark the mass ranges adopted in CAT for light and heavy BH seeds formation. 
     The candidate AGN in CEERS-1019 identified by \citet{larson2023} is marked with a black star at $z = 8.7$ ($t_{\rm H} = 0.570$ Gyr), the candidate AGN in UHZ1 at $z \simeq 10.3$ ($t_{\rm H} = 0.46$ Gyr) reported by \citet{bogdan2023} is shown by a smaller black star with errorbars. Finally, the AGN in GNz11 reported by \citet{maiolino2023a} is shown as a red star at $z=10.6$, ($t_{\rm H} = 0.44$ Gyr), and AGNs at lower redshift identified by Maiolino et al. (2023c) are shown by hollow red stars. 
     We also show the BH mass and redshift of systems identified in JWST surveys by \citet{kocevski2023} (yellow pentagons), \citet{ubler2023} (blue star), \citet{harikane2023} (green hollow stars), and the BH masses measured in quasars at $z \sim 6 - 7.5$ (lavender data points, \citealt{inayoshi2020}). Finally, in both panels, triangles connected with dotted lines indicate the maximum BH mass predicted in each model (yellow), and the minimum BH mass detectable in a survey like CEERS (turquoise) and JADES (violet). The vertical dotted lines show the corresponding maximum observable redshift in the two surveys. According to the EL model, systems like GNz11 and CEERS-1019 are among the most massive BHs at their corresponding redshift, and must originate from heavy black hole seeds formed at $14 \leq z \le 16$. In the SE model, they can originate from rapidly growing light seeds formed at $z \geq 20$ and/or from heavy BH seeds formed at $14 \leq z \le 16$, and do not represent the most extreme objects. In both models, we find that GNz11 may be a progenitor of CEERS-1019, if the latter hosts an AGN (see text), and that these BHs continue to grow, reaching a BH mass comparable to those measured in quasars at $6 \le z \le 7.5$. The estimated mass of the candidate AGN in UHZ1 exceeds the maximum BH mass predicted by EL model at $z = 10.3$, but is smaller than the maximum BH mass predicted by the SE.}
    \label{fig:BHtracks}
\end{figure*}

\subsection{Tracking the origin and fate of $z \simeq 9 - 11$ SMBHs} 
\label{sec:tracks}

The analysis of the emission properties of the best candidates of GNz11 and CEERS-1019 (among which we show in Figs. \ref{fig:ceersSED} and \ref{fig:gnz11SED} some representative cases) supports the interpretation that the central point source of GNz11 could be powered by a $1.5 \times 10^{6} \, \rm M_\odot$ super-Eddington accreting BH ($\lambda_{\rm Edd} \simeq 2 - 3$), while 
CEERS-1019 may harbour a more massive BH, with $M_{\rm BH} \gtrsim 10^7 M_\odot$, accreting sub-Eddington ($\lambda_{\rm Edd} \lesssim 0.5$), with a dominant emission from the host galaxy. 

Here we go one step further, and we investigate the origin and fate of these systems. In Fig. \ref{fig:BHtracks} we show their BH mass as a function of redshift (lower horizontal axis) and time (upper horizontal axis). Evolutionary tracks for the best candidates of CEERS-1019 at $z = 8.7$ (filled black star) in the EL model, and of GNz11 at $z = 10.6$ (filled red star) in the SE model are shown with blue and orange solid lines in the upper and lower panels, respectively. We also show with a smaller filled black star with errorbars the candidate system in UHZ1 at $z = 10.3$. The dark blue and red coloured points along the tracks mark the occurrence of BH mergers, while gray points indicate the mass of the initial BH seed and the mass at the observed redshift of GNz11 and CEERS-1019. We also show the BH masses and redshifts of AGNs identified by JWST at $z < 7$, using the same symbols and colours adopted in Figs. \ref{fig:mbhlbol} and \ref{fig:mbhmstar}. The lavender data points indicate the BHs measured in quasars at $z \sim 6 - 7.5$ \citep{inayoshi2020}. The gray horizontal bands identify the mass ranges corresponding to light and heavy BH seeds in CAT. In both panels, the dotted lines with triangles indicate the maximum BH mass that is predicted by CAT at $4 \leq z \leq 20$ (yellow), and the minimum BH mass that would be detectable in a survey like CEERS (turquoise) and JADES (violet). We also mark with vertical dotted lines the maximum redshift at which BHs would be observable in these two surveys, that correspond to the redshift at which the minimum detectable BH mass equals the maximum BH mass predicted by each model.

Our results suggest that there might be an evolutionary link between systems with a BH mass similar to GNz11 at $z = 10.6$, and systems with a mass similar to CEERS-1019 at $z = 8.7$. However, their formation history is very different in the two models: in the EL model, systems like GNz11 and CEERS-1019 originate from heavy BH seeds formed at $14 \leq z \le 16$, and are among the most massive BHs at their corresponding redshift. In the SE model, instead, all the tracks start at $z \gtrsim 20$ from light BH seeds, although mergers can occur with heavy BH seeds as they cross the heavy BH seed mass, at $z \le 16$, indicating that the genealogy of systems like GNz11 and CEERS-1019 could be characterized by both light and heavy BH seeds ancestors. This is consistent with the results presented in \citet{trinca2022} (see in particular their Fig. 7), which show that less than $\simeq 20\%$ of BHs with masses $\simeq 10^6 M_\odot - 10^7 M_\odot$ at $8 \leq z \leq 10$ descend from a heavy BH seed. We also note that in the SE model, the evolutionary tracks are always well below the maximum BH mass predicted by the model, meaning that systems similar to GNz11 and CEERS-1019 are not the most massive among the BH population at their corresponding redshift. Conversely, the candidate AGN at $z=10.3$ reported by \citet{bogdan2023}, if confirmed, would be among the most extreme BHs, in terms of BH mass, predicted by the SE models, and too massive to be formed in the EL model. Interestingly, at $z < 9 - 11$ the selected systems continue to grow in mass: in both models they could be the progenitors of SMBHs powering quasars at $6 \leq z \leq 7$, and of the most massive among the AGNs detected with JWST at $z < 7$. Conversely, JWST observed AGNs with mass $\lesssim 3 \times 10^7 M_\odot$ at $z < 7$ must originate from less efficiently growing seeds (see for instance the scenario proposed by \citealt{sassano2023}), and their evolution is therefore unrelated to GNz11-like or CEERS-1019-like systems.

If we compare the maximum BH mass predicted in each model with the minimum BH mass that would be observable by CEERS and JADES, we find that CEERS is not expected to observe accreting BHs beyond $z \simeq 9$ in the EL model, and CEERS-1019 represent the most massive BH at $z = 8.7$ and it is $\sim 0.6$ dex larger than the minimum detectable BH mass. JADES has the sensitivity to observe accreting BHs out to $z \simeq 12$, and GNz11 is $\sim 0.4$ dex smaller than the maximum BH mass at $z = 10.6$ and  $\sim 0.6$ dex larger than the minimum detectable BH mass. In the SE model these figures change considerably: we predict that CEERS would be able to detect BHs with masses $\gtrsim 3 \times 10^5 M_\odot - 10^6 M_\odot$ out to $z \simeq 15$, and JADES would be able to detect BHs with masses $\gtrsim 10^5 M_\odot - 3 \times 10^5 M_\odot$ out to $z \sim 18$. 

While BHs with masses comparable to those of GNz11 and CEERS-1019 can be explained in both models, the predicted discovery space is very different in the two scenarios: we can expect that JWST will continue to discover small but luminous AGNs at $z \lesssim 9 - 12$. However, if future surveys as sensitive as CEERS and JADES were to discover accreting BHs at $z > 12$, this would be a strong indication that early BH seeds must experience phases of super-Eddington growth.

\section{Discussion and Conclusions}
\label{sec:conclusions}

The recent discovery of more than 40 new AGNs at $z>4$ in ongoing surveys with JWST, including one system at $z = 10.6$, with estimated BH masses $\leq 10^7 M_\odot$, offers the unprecedented opportunity to constrain theories for the formation and evolution of SMBHs with masses $> 10^9 M_\odot$ powering quasars at $z > 6$. Given the high rate of AGNs discovered with JWST, two questions naturally arise:  are we surprised to find these accreting BHs in the nuclei of galaxies at $z \sim 5  - 11$? And can we use their estimated properties to constrain the nature and growth of the first BH seeds? 

Prompted by these questions, we use the Cosmic Archaeology Tool (CAT, T22) semi-analytical model to explore the population of BHs and their host galaxies at $5 \leq z \leq 11$, and compare their predicted properties to those estimated for the AGNs and host galaxies recently detected by JWST.

Following T22, we assume that light BH seeds form as remnants of Pop III stars and their masses range from 10s to 100s $M_\odot$, while heavy BH seeds form with a constant mass of $10^5 M_\odot$ from the collapse of supermassive stars. Starting from these two BH seeds flavours, we consider two alternative BH accretion models: in the first model, BH accretion is Eddington-limited (EL), while in the second model, light and heavy BH seeds can accrete at super-Eddington rates in short phases triggered by galaxy major mergers (SE). We note here that our description of super-Eddington accretion via a slim disk \citep{madau2014a, lupi2014, volonteri2015, pezzulli2016} is simplified (see the discussion in T22). High-resolution hydrodynamical simulations show that in this regime BH growth is very intermittent, due to radiative and mechanical feedback from the growing BH \citep{regan2019, massonneau2023}, and that efficient, super-Eddington growth may be possible onto heavy BH seeds \citep{inayoshi2022}, and less likely onto seeds with masses $\leq 10^3 M_\odot$ \citep{smith2018, sassano2023}. With this caveat in mind, in the present study we have considered both the EL and the SE models to explore how the assumed BH growth compares with current observational constraints provided by the recently JWST detected AGNs. 

We first consider the predicted distribution of nuclear BHs and their host galaxies in the BH mass - luminosity and BH - stellar mass planes at $5 \leq z \leq 11$, and investigate the properties of BHs that are luminous enough to be detectable by JWST in surveys like JADES and CEERS (see also \citealt{trinca2023bh}). We find that:

\begin{itemize}
\item In the EL model, BHs descending from light seeds grow very inefficiently, their masses and luminosities remain $\lesssim 10^4 M_\odot$ and ${\rm Log} (L_{\rm bol}) [\rm erg/s] \lesssim 39.5$, and are undermassive with respect to their host galaxies. All the systems detectable with JADES and CEERS appear to be descendants of heavy BH seeds. Even for these more efficiently growing seeds, their masses remain $\lesssim 10^6 M_\odot$ until their host galaxy has grown to a mass of at least $M_{\rm star} > 10^8 M_\odot$, above which the growth of the BHs adjust to $M_{\rm BH}/M_{\rm star} \lesssim 0.001$, with a large scatter. At all redshifts, most of the observable systems have $0.001 \lesssim M_{\rm BH}/M_{\rm star} \lesssim 0.01$. Interestingly, a small fraction of systems observable with JADES at $z \gtrsim 7$ is made by heavy BH seeds with $\lesssim 10^6 M_\odot$ which are over-massive with respect to their host galaxies, as expected if these systems keep memory of their peculiar birth conditions \citep{agarwal2013, natarajan2017, visbal2018, valiante2018observability}.

\item In the SE model, even light BH seeds grow more efficiently. However, there is a luminosity gap in the BH mass - luminosity plane, with two disjoint, parallel diagonal tracks, corresponding to $1 \lesssim \lambda_{\rm Edd} \lesssim 10$ and $\lambda_{\rm Edd} \lesssim 0.01 - 0.1$. This reflects the short duration of the SE phase, during which systems are shifted on the upper track, which becomes progressively less populated with decreasing redshift. The number of detectable systems is larger than in the EL model, particularly at $z \gtrsim 12$, where most of the systems are detectable when accreting at SE rates. This prediction could be test observationally with surveys like JADES and CEERS. At all redshifts, detectable BHs evolve with $M_{\rm BH}/M_{\rm star} \simeq 0.01 - 0.1$, and are overmassive with respect to the canonical value $M_{\rm BH}/M_{\rm star} \simeq 0.001$ \citep{reines2015, greene2020}.

\item  The comparison with JWST detected AGNs at $4 < z < 10.6$ shows that their estimated BH luminosities are better reproduced by the SE model, while their host galaxy masses are better matched by the EL model. In particular, we identify the subsample of detectable sources with BH masses that match the estimated values for GNz11 \citep{maiolino2023a} and CEERS-1019, if this system hosts an AGN \citep{larson2023}, and whose stellar mass (for sources in the EL model) or luminosity (for sources in the SE model) is consistent with the observed values. For these subsample of systems, which provide our best synthetic AGN analogues for GNz-11 and CEERS-1019, we compute their SEDs by coupling CAT predictions with the Cloudy photoionization code, and compare them with the observed JWST photometry. Our analysis supports the interpretation that the central point source of GNz11 could be powered by a $1.5 \times 10^{6} \, \rm M_\odot$ super-Eddington accreting BH ($\lambda_{\rm Edd} \simeq 2 - 3$), while the observed emission from CEERS-1019 is dominated by the host galaxy: if CEERS-1019  
harbours an accreting BH, we find it to have a mass of $M_{\rm BH} \simeq 10^7 M_\odot$, and to be accreting at sub-Eddington rates ($\lambda_{\rm Edd} \simeq 0.5$). Yet, these conditions are met in only 10\% of the systems predicted by CAT EL model at $z \simeq 9$, when these are selected to have stellar masses and star formation rates consistent with the values inferred for CEERS-1019 \citep{larson2023}. In the rest of the EL sample, our synthetic counterparts of CEERS-1019 host smaller (${\rm log}(M_{\rm BH}/M_\odot) = 5.25 - 6.5$), and less active BHs ($\lambda_{\rm Edd} = 0.03 - 0.3$), or - in the SE model - more massive (${\rm log}(M_{\rm BH}/M_\odot) = 7.4 - 8.5$) and less active BHs ($\lambda_{\rm Edd} = 0.009 - 0.03$).

\item We also analyse the evolutionary tracks of CEERS-1019 and GNz11 synthetic analogues in the EL and in the SE models to investigate the origin and fate of these systems. We find that there might be an evolutionary link between systems with a BH mass similar to GNz11 at $z = 10.6$ and systems with a BH mass at $z = 8.7$ similar to CEERS-1019, if this is an AGN. However, their formation history is very different in the two models: in the EL model, systems like GNz11 and CEERS-1019 originate from heavy BH seeds formed at $14 \leq z \le 16$, and are among the most massive BHs at their corresponding redshift. In the SE model, instead, systems similar to GNz11 and CEERS-1019 are not the most massive among the BH population at their corresponding redshift; their evolutionary tracks start at $z \gtrsim 20$ from light BH seeds, although mergers can occur with heavy BH seeds at $z \le 16$, indicating that they could descend from both light and heavy BH seeds ancestors.  In both models GNz11-like or CEERS-1019-like systems continue to grow at $z < 9 - 11$, reaching masses comparable to the SMBHs powering quasars at $6 \leq z \leq 7$, and of the most massive among the AGNs detected with JWST at $z < 7$. Conversely, the sample of JWST AGNs with mass $\lesssim 3 \times 10^7 M_\odot$ at $z < 7$ must originate from less efficiently growing seeds.
\end{itemize}

Our results suggest that the AGNs discovered in ongoing JWST surveys are providing an amazing confirmation of the rich BH landscape at cosmic dawn predicted by theoretical models. AGNs with properties similar to those estimated for JWST detected AGNs, including the most distant source, GNz11, and CEERS-1019, if this hosts an AGN, can be explained in scenarios where BHs originate from Eddington-limited gas accretion onto heavy BH seeds, or if light and heavy BH seeds can experience short phases of super-Eddington accretion. We expect that surveys like CEERS and JADES will continue to discover small but luminous AGNs at $z \lesssim 9 - 12$ \citep{trinca2023bh}. However, if JWST surveys as sensitive as CEERS and JADES were to discover accreting BHs at $z > 12$, and/or the BH mass estimated for the AGN candidate hosted in UHZ1 with a photometric redshift of $z = 10.3$ were to be confirmed with future observations, we would have a strong indication that conditions enabling super-Eddington accretion, such as a high optical depth and high gas supply rates from large scales \citep{mayer2019b}, should be naturally met and that feedback in the super-Eddington phase does not curtail significantly its duration.

\section*{Acknowledgements}
We are very grateful to the Referee, Mike Strauss, for his very constructive and insightful comments. 
RS and LG acknowledge support from the Amaldi Research
Center funded by the MIUR program “Dipartimento di Eccellenza ”(CUP:B81I18001170001) and from the INFN TEONGRAV specific initiative.

\section*{Data Availability}
The simulated data underlying this article will be shared on reasonable request to the corresponding author.


\bibliographystyle{mnras}
\bibliography{CAT_firstBHtracks}

\begin{thebibliography}{}
\makeatletter
\relax
\def\mn@urlcharsother{\let\do\@makeother \do\$\do\&\do\#\do\^\do\_\do\%\do\~}
\def\mn@doi{\begingroup\mn@urlcharsother \@ifnextchar [ {\mn@doi@}
  {\mn@doi@[]}}
\def\mn@doi@[#1]#2{\def\@tempa{#1}\ifx\@tempa\@empty \href
  {http://dx.doi.org/#2} {doi:#2}\else \href {http://dx.doi.org/#2} {#1}\fi
  \endgroup}
\def\mn@eprint#1#2{\mn@eprint@#1:#2::\@nil}
\def\mn@eprint@arXiv#1{\href {http://arxiv.org/abs/#1} {{\tt arXiv:#1}}}
\def\mn@eprint@dblp#1{\href {http://dblp.uni-trier.de/rec/bibtex/#1.xml}
  {dblp:#1}}
\def\mn@eprint@#1:#2:#3:#4\@nil{\def\@tempa {#1}\def\@tempb {#2}\def\@tempc
  {#3}\ifx \@tempc \@empty \let \@tempc \@tempb \let \@tempb \@tempa \fi \ifx
  \@tempb \@empty \def\@tempb {arXiv}\fi \@ifundefined
  {mn@eprint@\@tempb}{\@tempb:\@tempc}{\expandafter \expandafter \csname
  mn@eprint@\@tempb\endcsname \expandafter{\@tempc}}}

\bibitem[\protect\citeauthoryear{{Adams} et~al.,}{{Adams}
  et~al.}{2023}]{Adams2023}
{Adams} N.~J.,  et~al., 2023, \mn@doi [\mnras]
  {10.1093/mnras/stac334710.48550/arXiv.2207.11217}, \href
  {https://ui.adsabs.harvard.edu/abs/2023MNRAS.518.4755A} {518, 4755}

\bibitem[\protect\citeauthoryear{{Agarwal}, {Davis}, {Khochfar}, {Natarajan}
  \& {Dunlop}}{{Agarwal} et~al.}{2013}]{agarwal2013}
{Agarwal} B.,  {Davis} A.~J.,  {Khochfar} S.,  {Natarajan} P.,   {Dunlop}
  J.~S.,  2013, \mn@doi [\mnras] {10.1093/mnras/stt696}, \href
  {https://ui.adsabs.harvard.edu/abs/2013MNRAS.432.3438A} {432, 3438}

\bibitem[\protect\citeauthoryear{{Angl{\'e}s-Alc{\'a}zar},
  {Faucher-Gigu{\`e}re}, {Quataert}, {Hopkins}, {Feldmann}, {Torrey}, {Wetzel}
  \& {Kere{\v{s}}}}{{Angl{\'e}s-Alc{\'a}zar} et~al.}{2017}]{alcazar2017}
{Angl{\'e}s-Alc{\'a}zar} D.,  {Faucher-Gigu{\`e}re} C.-A.,  {Quataert} E.,
  {Hopkins} P.~F.,  {Feldmann} R.,  {Torrey} P.,  {Wetzel} A.,   {Kere{\v{s}}}
  D.,  2017, \mn@doi [\mnras] {10.1093/mnrasl/slx161}, \href
  {https://ui.adsabs.harvard.edu/abs/2017MNRAS.472L.109A} {472, L109}

\bibitem[\protect\citeauthoryear{{Arrabal Haro} et~al.,}{{Arrabal Haro}
  et~al.}{2023}]{arrabalharo2023}
{Arrabal Haro} P.,  et~al., 2023, arXiv e-prints, \href
  {https://ui.adsabs.harvard.edu/abs/2023arXiv230315431A} {p. arXiv:2303.15431}

\bibitem[\protect\citeauthoryear{{Backhaus} et~al.,}{{Backhaus}
  et~al.}{2022}]{backhaus2022}
{Backhaus} B.~E.,  et~al., 2022, \mn@doi [\apj] {10.3847/1538-4357/ac3919},
  \href {https://ui.adsabs.harvard.edu/abs/2022ApJ...926..161B} {926, 161}

\bibitem[\protect\citeauthoryear{{Baldwin}, {Phillips}  \&
  {Terlevich}}{{Baldwin} et~al.}{1981}]{baldwin1981}
{Baldwin} J.~A.,  {Phillips} M.~M.,   {Terlevich} R.,  1981, \mn@doi [\pasp]
  {10.1086/130766}, \href
  {https://ui.adsabs.harvard.edu/abs/1981PASP...93....5B} {93, 5}

\bibitem[\protect\citeauthoryear{Ba{\~n}ados et~al.,}{Ba{\~n}ados
  et~al.}{2018}]{banados2018}
Ba{\~n}ados E.,  et~al., 2018, Nature, 553, 473

\bibitem[\protect\citeauthoryear{{Barrow}, {Aykutalp}  \& {Wise}}{{Barrow}
  et~al.}{2018}]{barrow2018}
{Barrow} K. S.~S.,  {Aykutalp} A.,   {Wise} J.~H.,  2018, \mn@doi [Nature
  Astronomy] {10.1038/s41550-018-0569-y}, \href
  {https://ui.adsabs.harvard.edu/abs/2018NatAs...2..987B} {2, 987}

\bibitem[\protect\citeauthoryear{{Barrufet} et~al.,}{{Barrufet}
  et~al.}{2022}]{barrufet2022}
{Barrufet} L.,  et~al., 2022, \mn@doi [arXiv e-prints]
  {10.48550/arXiv.2207.14733}, \href
  {https://ui.adsabs.harvard.edu/abs/2022arXiv220714733B} {p. arXiv:2207.14733}

\bibitem[\protect\citeauthoryear{{Bogdan} et~al.,}{{Bogdan}
  et~al.}{2023}]{bogdan2023}
{Bogdan} A.,  et~al., 2023, \mn@doi [arXiv e-prints]
  {10.48550/arXiv.2305.15458}, \href
  {https://ui.adsabs.harvard.edu/abs/2023arXiv230515458B} {p. arXiv:2305.15458}

\bibitem[\protect\citeauthoryear{{Bortolas}, {Capelo}, {Zana}, {Mayer},
  {Bonetti}, {Dotti}, {Davies}  \& {Madau}}{{Bortolas}
  et~al.}{2020}]{bortolas2020}
{Bortolas} E.,  {Capelo} P.~R.,  {Zana} T.,  {Mayer} L.,  {Bonetti} M.,
  {Dotti} M.,  {Davies} M.~B.,   {Madau} P.,  2020, arXiv e-prints, \href
  {https://ui.adsabs.harvard.edu/abs/2020arXiv200502409B} {p. arXiv:2005.02409}

\bibitem[\protect\citeauthoryear{{Bunker} et~al.,}{{Bunker}
  et~al.}{2023}]{bunker2023}
{Bunker} A.~J.,  et~al., 2023, \mn@doi [arXiv e-prints]
  {10.48550/arXiv.2302.07256}, \href
  {https://ui.adsabs.harvard.edu/abs/2023arXiv230207256B} {p. arXiv:2302.07256}

\bibitem[\protect\citeauthoryear{{Cameron} et~al.,}{{Cameron}
  et~al.}{2023}]{cameron2023}
{Cameron} A.~J.,  et~al., 2023, \mn@doi [arXiv e-prints]
  {10.48550/arXiv.2302.04298}, \href
  {https://ui.adsabs.harvard.edu/abs/2023arXiv230204298C} {p. arXiv:2302.04298}

\bibitem[\protect\citeauthoryear{{Carnall} et~al.,}{{Carnall}
  et~al.}{2023}]{carnall2023}
{Carnall} A.~C.,  et~al., 2023, \mn@doi [\mnras] {10.1093/mnras/stad369}, \href
  {https://ui.adsabs.harvard.edu/abs/2023MNRAS.520.3974C} {520, 3974}

\bibitem[\protect\citeauthoryear{{Castellano} et~al.,}{{Castellano}
  et~al.}{2022}]{castellano2022}
{Castellano} M.,  et~al., 2022, \mn@doi [\apjl] {10.3847/2041-8213/ac94d0},
  \href {https://ui.adsabs.harvard.edu/abs/2022ApJ...938L..15C} {938, L15}

\bibitem[\protect\citeauthoryear{Chon \& Omukai}{Chon \&
  Omukai}{2020}]{chon2020}
Chon S.,  Omukai K.,  2020, Supermassive Star Formation via Super Competitive
  Accretion in Slightly Metal-enriched Clouds (\mn@eprint {arXiv} {2001.06491})

\bibitem[\protect\citeauthoryear{{Cole}, {Lacey}, {Baugh}  \& {Frenk}}{{Cole}
  et~al.}{2000}]{cole2000}
{Cole} S.,  {Lacey} C.~G.,  {Baugh} C.~M.,   {Frenk} C.~S.,  2000, \mn@doi
  [\mnras] {10.1046/j.1365-8711.2000.03879.x}, \href
  {https://ui.adsabs.harvard.edu/abs/2000MNRAS.319..168C} {319, 168}

\bibitem[\protect\citeauthoryear{{Curti} et~al.,}{{Curti}
  et~al.}{2023}]{curti2023}
{Curti} M.,  et~al., 2023, \mn@doi [\mnras] {10.1093/mnras/stac2737}, \href
  {https://ui.adsabs.harvard.edu/abs/2023MNRAS.518..425C} {518, 425}

\bibitem[\protect\citeauthoryear{{Curtis-Lake} et~al.,}{{Curtis-Lake}
  et~al.}{2022}]{curtislake2022}
{Curtis-Lake} E.,  et~al., 2022, \mn@doi [arXiv e-prints]
  {10.48550/arXiv.2212.04568}, \href
  {https://ui.adsabs.harvard.edu/abs/2022arXiv221204568C} {p. arXiv:2212.04568}

\bibitem[\protect\citeauthoryear{Devecchi \& Volonteri}{Devecchi \&
  Volonteri}{2009}]{devecchi2009}
Devecchi B.,  Volonteri M.,  2009, The Astrophysical Journal, 694, 302

\bibitem[\protect\citeauthoryear{{Devecchi}, {Volonteri}, {Rossi}, {Colpi}  \&
  {Portegies Zwart}}{{Devecchi} et~al.}{2012}]{devecchi2012}
{Devecchi} B.,  {Volonteri} M.,  {Rossi} E.~M.,  {Colpi} M.,   {Portegies
  Zwart} S.,  2012, \mn@doi [\mnras] {10.1111/j.1365-2966.2012.20406.x}, \href
  {https://ui.adsabs.harvard.edu/abs/2012MNRAS.421.1465D} {421, 1465}

\bibitem[\protect\citeauthoryear{{Ding} et~al.,}{{Ding}
  et~al.}{2022}]{ding2022}
{Ding} X.,  et~al., 2022, \mn@doi [arXiv e-prints] {10.48550/arXiv.2211.14329},
  \href {https://ui.adsabs.harvard.edu/abs/2022arXiv221114329D} {p.
  arXiv:2211.14329}

\bibitem[\protect\citeauthoryear{{Donnan} et~al.,}{{Donnan}
  et~al.}{2023}]{Donnan2023}
{Donnan} C.~T.,  et~al., 2023, \mn@doi [\mnras]
  {10.1093/mnras/stac347210.48550/arXiv.2207.12356}, \href
  {https://ui.adsabs.harvard.edu/abs/2023MNRAS.518.6011D} {518, 6011}

\bibitem[\protect\citeauthoryear{{Dopita} et~al.,}{{Dopita}
  et~al.}{2006}]{dopita2006}
{Dopita} M.~A.,  et~al., 2006, \mn@doi [\apjs] {10.1086/508261}, \href
  {https://ui.adsabs.harvard.edu/abs/2006ApJS..167..177D} {167, 177}

\bibitem[\protect\citeauthoryear{{Endsley}, {Stark}, {Whitler}, {Topping},
  {Chen}, {Plat}, {Chisholm}  \& {Charlot}}{{Endsley}
  et~al.}{2023}]{endsley2023}
{Endsley} R.,  {Stark} D.~P.,  {Whitler} L.,  {Topping} M.~W.,  {Chen} Z.,
  {Plat} A.,  {Chisholm} J.,   {Charlot} S.,  2023, \mn@doi [\mnras]
  {10.1093/mnras/stad1919}, \href
  {https://ui.adsabs.harvard.edu/abs/2023MNRAS.tmp.1872E} {}

\bibitem[\protect\citeauthoryear{{Fan}, {Banados}  \& {Simcoe}}{{Fan}
  et~al.}{2022}]{fan2022}
{Fan} X.,  {Banados} E.,   {Simcoe} R.~A.,  2022, \mn@doi [arXiv e-prints]
  {10.48550/arXiv.2212.06907}, \href
  {https://ui.adsabs.harvard.edu/abs/2022arXiv221206907F} {p. arXiv:2212.06907}

\bibitem[\protect\citeauthoryear{{Ferland} et~al.,}{{Ferland}
  et~al.}{2017}]{ferland2017}
{Ferland} G.~J.,  et~al., 2017, \rmxaa, \href
  {https://ui.adsabs.harvard.edu/abs/2017RMxAA..53..385F} {53, 385}

\bibitem[\protect\citeauthoryear{{Ferrara}, {Salvadori}, {Yue}  \&
  {Schleicher}}{{Ferrara} et~al.}{2014}]{ferrara2014}
{Ferrara} A.,  {Salvadori} S.,  {Yue} B.,   {Schleicher} D.,  2014, \mn@doi
  [\mnras] {10.1093/mnras/stu1280}, \href
  {https://ui.adsabs.harvard.edu/abs/2014MNRAS.443.2410F} {443, 2410}

\bibitem[\protect\citeauthoryear{{Fioc} \& {Rocca-Volmerange}}{{Fioc} \&
  {Rocca-Volmerange}}{2019}]{fioc2019}
{Fioc} M.,  {Rocca-Volmerange} B.,  2019, \mn@doi [\aap]
  {10.1051/0004-6361/201833556}, \href
  {https://ui.adsabs.harvard.edu/abs/2019A&A...623A.143F} {623, A143}

\bibitem[\protect\citeauthoryear{{Fujimoto} et~al.,}{{Fujimoto}
  et~al.}{2022}]{fujimoto2022}
{Fujimoto} S.,  et~al., 2022, \mn@doi [\nat] {10.1038/s41586-022-04454-1},
  \href {https://ui.adsabs.harvard.edu/abs/2022Natur.604..261F} {604, 261}

\bibitem[\protect\citeauthoryear{{Furtak} et~al.,}{{Furtak}
  et~al.}{2022}]{furtak2022}
{Furtak} L.~J.,  et~al., 2022, \mn@doi [arXiv e-prints]
  {10.48550/arXiv.2212.10531}, \href
  {https://ui.adsabs.harvard.edu/abs/2022arXiv221210531F} {p. arXiv:2212.10531}

\bibitem[\protect\citeauthoryear{{Goulding} \& {Greene}}{{Goulding} \&
  {Greene}}{2022}]{goulding2022}
{Goulding} A.~D.,  {Greene} J.~E.,  2022, arXiv e-prints, \href
  {https://ui.adsabs.harvard.edu/abs/2022arXiv220802822G} {p. arXiv:2208.02822}

\bibitem[\protect\citeauthoryear{{Greene}, {Strader}  \& {Ho}}{{Greene}
  et~al.}{2020}]{greene2020}
{Greene} J.~E.,  {Strader} J.,   {Ho} L.~C.,  2020, \mn@doi [\araa]
  {10.1146/annurev-astro-032620-021835}, \href
  {https://ui.adsabs.harvard.edu/abs/2020ARA&A..58..257G} {58, 257}

\bibitem[\protect\citeauthoryear{{Gutkin}, {Charlot}  \& {Bruzual}}{{Gutkin}
  et~al.}{2016}]{gutkin2016}
{Gutkin} J.,  {Charlot} S.,   {Bruzual} G.,  2016, \mn@doi [\mnras]
  {10.1093/mnras/stw1716}, \href
  {https://ui.adsabs.harvard.edu/abs/2016MNRAS.462.1757G} {462, 1757}

\bibitem[\protect\citeauthoryear{{Habouzit}, {Volonteri}  \&
  {Dubois}}{{Habouzit} et~al.}{2017}]{habouzit2017}
{Habouzit} M.,  {Volonteri} M.,   {Dubois} Y.,  2017, \mn@doi [\mnras]
  {10.1093/mnras/stx666}, \href
  {https://ui.adsabs.harvard.edu/abs/2017MNRAS.468.3935H} {468, 3935}

\bibitem[\protect\citeauthoryear{{Harikane} et~al.,}{{Harikane}
  et~al.}{2022}]{harikane2022b}
{Harikane} Y.,  et~al., 2022, \mn@doi [\apjs]
  {10.3847/1538-4365/ac3dfc10.48550/arXiv.2108.01090}, \href
  {https://ui.adsabs.harvard.edu/abs/2022ApJS..259...20H} {259, 20}

\bibitem[\protect\citeauthoryear{{Harikane} et~al.,}{{Harikane}
  et~al.}{2023}]{harikane2023}
{Harikane} Y.,  et~al., 2023, \mn@doi [arXiv e-prints]
  {10.48550/arXiv.2303.11946}, \href
  {https://ui.adsabs.harvard.edu/abs/2023arXiv230311946H} {p. arXiv:2303.11946}

\bibitem[\protect\citeauthoryear{{Inayoshi}, {Visbal}  \& {Haiman}}{{Inayoshi}
  et~al.}{2020}]{inayoshi2020}
{Inayoshi} K.,  {Visbal} E.,   {Haiman} Z.,  2020, \mn@doi [\araa]
  {10.1146/annurev-astro-120419-014455}, \href
  {https://ui.adsabs.harvard.edu/abs/2020ARA&A..58...27I} {58, 27}

\bibitem[\protect\citeauthoryear{{Inayoshi}, {Onoue}, {Sugahara}, {Inoue}  \&
  {Ho}}{{Inayoshi} et~al.}{2022}]{inayoshi2022}
{Inayoshi} K.,  {Onoue} M.,  {Sugahara} Y.,  {Inoue} A.~K.,   {Ho} L.~C.,
  2022, \mn@doi [\apjl] {10.3847/2041-8213/ac6f01}, \href
  {https://ui.adsabs.harvard.edu/abs/2022ApJ...931L..25I} {931, L25}

\bibitem[\protect\citeauthoryear{{Kashino}, {Lilly}, {Matthee}, {Eilers},
  {Mackenzie}, {Bordoloi}  \& {Simcoe}}{{Kashino} et~al.}{2023}]{kashino2023}
{Kashino} D.,  {Lilly} S.~J.,  {Matthee} J.,  {Eilers} A.-C.,  {Mackenzie} R.,
  {Bordoloi} R.,   {Simcoe} R.~A.,  2023, \mn@doi [\apj]
  {10.3847/1538-4357/acc588}, \href
  {https://ui.adsabs.harvard.edu/abs/2023ApJ...950...66K} {950, 66}

\bibitem[\protect\citeauthoryear{{Kocevski} et~al.,}{{Kocevski}
  et~al.}{2023}]{kocevski2023}
{Kocevski} D.~D.,  et~al., 2023, \mn@doi [arXiv e-prints]
  {10.48550/arXiv.2302.00012}, \href
  {https://ui.adsabs.harvard.edu/abs/2023arXiv230200012K} {p. arXiv:2302.00012}

\bibitem[\protect\citeauthoryear{{Labbe} et~al.,}{{Labbe}
  et~al.}{2022}]{labbe2022}
{Labbe} I.,  et~al., 2022, \mn@doi [arXiv e-prints]
  {10.48550/arXiv.2207.12446}, \href
  {https://ui.adsabs.harvard.edu/abs/2022arXiv220712446L} {p. arXiv:2207.12446}

\bibitem[\protect\citeauthoryear{{Labbe} et~al.,}{{Labbe}
  et~al.}{2023}]{labbe2023}
{Labbe} I.,  et~al., 2023, \mn@doi [arXiv e-prints]
  {10.48550/arXiv.2306.07320}, \href
  {https://ui.adsabs.harvard.edu/abs/2023arXiv230607320L} {p. arXiv:2306.07320}

\bibitem[\protect\citeauthoryear{{Larson} et~al.,}{{Larson}
  et~al.}{2023}]{larson2023}
{Larson} R.~L.,  et~al., 2023, \mn@doi [arXiv e-prints]
  {10.48550/arXiv.2303.08918}, \href
  {https://ui.adsabs.harvard.edu/abs/2023arXiv230308918L} {p. arXiv:2303.08918}

\bibitem[\protect\citeauthoryear{{Looser} et~al.,}{{Looser}
  et~al.}{2023}]{looser2023}
{Looser} T.~J.,  et~al., 2023, \mn@doi [arXiv e-prints]
  {10.48550/arXiv.2306.02470}, \href
  {https://ui.adsabs.harvard.edu/abs/2023arXiv230602470L} {p. arXiv:2306.02470}

\bibitem[\protect\citeauthoryear{{Lupi}, {Colpi}, {Devecchi}, {Galanti}  \&
  {Volonteri}}{{Lupi} et~al.}{2014}]{lupi2014}
{Lupi} A.,  {Colpi} M.,  {Devecchi} B.,  {Galanti} G.,   {Volonteri} M.,  2014,
  \mn@doi [\mnras] {10.1093/mnras/stu1120}, \href
  {https://ui.adsabs.harvard.edu/abs/2014MNRAS.442.3616L} {442, 3616}

\bibitem[\protect\citeauthoryear{{Lupi}, {Volonteri}, {Decarli}, {Bovino},
  {Silk}  \& {Bergeron}}{{Lupi} et~al.}{2019}]{lupi2019}
{Lupi} A.,  {Volonteri} M.,  {Decarli} R.,  {Bovino} S.,  {Silk} J.,
  {Bergeron} J.,  2019, \mn@doi [\mnras] {10.1093/mnras/stz1959}, \href
  {https://ui.adsabs.harvard.edu/abs/2019MNRAS.488.4004L} {488, 4004}

\bibitem[\protect\citeauthoryear{{Madau}, {Haardt}  \& {Dotti}}{{Madau}
  et~al.}{2014}]{madau2014a}
{Madau} P.,  {Haardt} F.,   {Dotti} M.,  2014, \mn@doi [\apjl]
  {10.1088/2041-8205/784/2/L38}, \href
  {https://ui.adsabs.harvard.edu/abs/2014ApJ...784L..38M} {784, L38}

\bibitem[\protect\citeauthoryear{{Maiolino} et~al.,}{{Maiolino}
  et~al.}{2023a}]{maiolino2023a}
{Maiolino} R.,  et~al., 2023a, \mn@doi [arXiv e-prints]
  {10.48550/arXiv.2305.12492}, \href
  {https://ui.adsabs.harvard.edu/abs/2023arXiv230512492M} {p. arXiv:2305.12492}

\bibitem[\protect\citeauthoryear{{Maiolino} et~al.,}{{Maiolino}
  et~al.}{2023b}]{maiolino2023b}
{Maiolino} R.,  et~al., 2023b, \mn@doi [arXiv e-prints]
  {10.48550/arXiv.2306.00953}, \href
  {https://ui.adsabs.harvard.edu/abs/2023arXiv230600953M} {p. arXiv:2306.00953}

\bibitem[\protect\citeauthoryear{{Martin} \& {Rouleau}}{{Martin} \&
  {Rouleau}}{1991}]{martin1991}
{Martin} P.~G.,  {Rouleau} F.,  1991, in {Malina} R.~F.,  {Bowyer} S.,  eds,
  Extreme Ultraviolet Astronomy. p.~341

\bibitem[\protect\citeauthoryear{{Massonneau}, {Dubois}, {Volonteri}  \&
  {Beckmann}}{{Massonneau} et~al.}{2023}]{massonneau2023}
{Massonneau} W.,  {Dubois} Y.,  {Volonteri} M.,   {Beckmann} R.~S.,  2023,
  \mn@doi [\aap] {10.1051/0004-6361/202244874}, \href
  {https://ui.adsabs.harvard.edu/abs/2023A&A...669A.143M} {669, A143}

\bibitem[\protect\citeauthoryear{{Mathis}, {Rumpl}  \& {Nordsieck}}{{Mathis}
  et~al.}{1977}]{mathis1977}
{Mathis} J.~S.,  {Rumpl} W.,   {Nordsieck} K.~H.,  1977, \mn@doi [\apj]
  {10.1086/155591}, \href
  {https://ui.adsabs.harvard.edu/abs/1977ApJ...217..425M} {217, 425}

\bibitem[\protect\citeauthoryear{{Matthee} et~al.,}{{Matthee}
  et~al.}{2023}]{matthee2023}
{Matthee} J.,  et~al., 2023, \mn@doi [arXiv e-prints]
  {10.48550/arXiv.2306.05448}, \href
  {https://ui.adsabs.harvard.edu/abs/2023arXiv230605448M} {p. arXiv:2306.05448}

\bibitem[\protect\citeauthoryear{{Mayer}}{{Mayer}}{2019}]{mayer2019b}
{Mayer} L.,  2019, in {Latif} M.,  {Schleicher} D.,  eds, , Formation of the
  First Black Holes.
pp 195--222, \mn@doi{10.1142/9789813227958_0011}

\bibitem[\protect\citeauthoryear{{Naidu} et~al.,}{{Naidu}
  et~al.}{2022a}]{naidu2022b}
{Naidu} R.~P.,  et~al., 2022a, \mn@doi [arXiv e-prints]
  {10.48550/arXiv.2208.02794}, \href
  {https://ui.adsabs.harvard.edu/abs/2022arXiv220802794N} {p. arXiv:2208.02794}

\bibitem[\protect\citeauthoryear{{Naidu} et~al.,}{{Naidu}
  et~al.}{2022b}]{naidu2022}
{Naidu} R.~P.,  et~al., 2022b, \mn@doi [\apjl]
  {10.3847/2041-8213/ac9b2210.48550/arXiv.2207.09434}, \href
  {https://ui.adsabs.harvard.edu/abs/2022ApJ...940L..14N} {940, L14}

\bibitem[\protect\citeauthoryear{{Nakajima} \& {Maiolino}}{{Nakajima} \&
  {Maiolino}}{2022}]{nakajima2022}
{Nakajima} K.,  {Maiolino} R.,  2022, \mn@doi [\mnras]
  {10.1093/mnras/stac1242}, \href
  {https://ui.adsabs.harvard.edu/abs/2022MNRAS.513.5134N} {513, 5134}

\bibitem[\protect\citeauthoryear{{Nakajima} et~al.,}{{Nakajima}
  et~al.}{2018}]{nakajima2018}
{Nakajima} K.,  et~al., 2018, \mn@doi [\aap] {10.1051/0004-6361/201731935},
  \href {https://ui.adsabs.harvard.edu/abs/2018A&A...612A..94N} {612, A94}

\bibitem[\protect\citeauthoryear{{Nakajima}, {Ouchi}, {Isobe}, {Harikane},
  {Zhang}, {Ono}, {Umeda}  \& {Oguri}}{{Nakajima} et~al.}{2023}]{nakajima2023}
{Nakajima} K.,  {Ouchi} M.,  {Isobe} Y.,  {Harikane} Y.,  {Zhang} Y.,  {Ono}
  Y.,  {Umeda} H.,   {Oguri} M.,  2023, \mn@doi [arXiv e-prints]
  {10.48550/arXiv.2301.12825}, \href
  {https://ui.adsabs.harvard.edu/abs/2023arXiv230112825N} {p. arXiv:2301.12825}

\bibitem[\protect\citeauthoryear{Natarajan, Pacucci, Ferrara, Agarwal, Ricarte,
  Zackrisson  \& Cappelluti}{Natarajan et~al.}{2017a}]{natarajan2017unveiling}
Natarajan P.,  Pacucci F.,  Ferrara A.,  Agarwal B.,  Ricarte A.,  Zackrisson
  E.,   Cappelluti N.,  2017a, The Astrophysical Journal, 838, 117

\bibitem[\protect\citeauthoryear{{Natarajan}, {Pacucci}, {Ferrara}, {Agarwal},
  {Ricarte}, {Zackrisson}  \& {Cappelluti}}{{Natarajan}
  et~al.}{2017b}]{natarajan2017}
{Natarajan} P.,  {Pacucci} F.,  {Ferrara} A.,  {Agarwal} B.,  {Ricarte} A.,
  {Zackrisson} E.,   {Cappelluti} N.,  2017b, \mn@doi [\apj]
  {10.3847/1538-4357/aa6330}, \href
  {https://ui.adsabs.harvard.edu/abs/2017ApJ...838..117N} {838, 117}

\bibitem[\protect\citeauthoryear{{Oesch} et~al.,}{{Oesch}
  et~al.}{2023}]{oesch2023}
{Oesch} P.~A.,  et~al., 2023, \mn@doi [arXiv e-prints]
  {10.48550/arXiv.2304.02026}, \href
  {https://ui.adsabs.harvard.edu/abs/2023arXiv230402026O} {p. arXiv:2304.02026}

\bibitem[\protect\citeauthoryear{Omukai, Schneider  \& Haiman}{Omukai
  et~al.}{2008}]{omukai2008}
Omukai K.,  Schneider R.,   Haiman Z.,  2008, The Astrophysical Journal, 686,
  801

\bibitem[\protect\citeauthoryear{{Onoue} et~al.,}{{Onoue}
  et~al.}{2023}]{onoue2023}
{Onoue} M.,  et~al., 2023, \mn@doi [\apjl] {10.3847/2041-8213/aca9d3}, \href
  {https://ui.adsabs.harvard.edu/abs/2023ApJ...942L..17O} {942, L17}

\bibitem[\protect\citeauthoryear{{Parkinson}, {Cole}  \& {Helly}}{{Parkinson}
  et~al.}{2008}]{parkinson2008}
{Parkinson} H.,  {Cole} S.,   {Helly} J.,  2008, \mn@doi [\mnras]
  {10.1111/j.1365-2966.2007.12517.x}, \href
  {https://ui.adsabs.harvard.edu/abs/2008MNRAS.383..557P} {383, 557}

\bibitem[\protect\citeauthoryear{{Pezzulli}, {Valiante}  \&
  {Schneider}}{{Pezzulli} et~al.}{2016}]{pezzulli2016}
{Pezzulli} E.,  {Valiante} R.,   {Schneider} R.,  2016, \mn@doi [\mnras]
  {10.1093/mnras/stw505}, \href
  {https://ui.adsabs.harvard.edu/abs/2016MNRAS.458.3047P} {458, 3047}

\bibitem[\protect\citeauthoryear{{Planck Collaboration} et~al.,}{{Planck
  Collaboration} et~al.}{2018}]{planck2018}
{Planck Collaboration} et~al., 2018, arXiv e-prints, \href
  {https://ui.adsabs.harvard.edu/abs/2018arXiv180706209P} {p. arXiv:1807.06209}

\bibitem[\protect\citeauthoryear{{Regan}, {Downes}, {Volonteri}, {Beckmann},
  {Lupi}, {Trebitsch}  \& {Dubois}}{{Regan} et~al.}{2019}]{regan2019}
{Regan} J.~A.,  {Downes} T.~P.,  {Volonteri} M.,  {Beckmann} R.,  {Lupi} A.,
  {Trebitsch} M.,   {Dubois} Y.,  2019, \mn@doi [\mnras]
  {10.1093/mnras/stz1045}, \href
  {https://ui.adsabs.harvard.edu/abs/2019MNRAS.486.3892R} {486, 3892}

\bibitem[\protect\citeauthoryear{{Regan}, {Haiman}, {Wise}, {O'Shea}  \&
  {Norman}}{{Regan} et~al.}{2020}]{regan2020}
{Regan} J.~A.,  {Haiman} Z.,  {Wise} J.~H.,  {O'Shea} B.~W.,   {Norman} M.~L.,
  2020, arXiv e-prints, \href
  {https://ui.adsabs.harvard.edu/abs/2020arXiv200614625R} {p. arXiv:2006.14625}

\bibitem[\protect\citeauthoryear{{Reines} \& {Volonteri}}{{Reines} \&
  {Volonteri}}{2015}]{reines2015}
{Reines} A.~E.,  {Volonteri} M.,  2015, \mn@doi [\apj]
  {10.1088/0004-637X/813/2/82}, \href
  {https://ui.adsabs.harvard.edu/abs/2015ApJ...813...82R} {813, 82}

\bibitem[\protect\citeauthoryear{{Ricarte} \& {Natarajan}}{{Ricarte} \&
  {Natarajan}}{2018}]{ricarte2018b}
{Ricarte} A.,  {Natarajan} P.,  2018, \mn@doi [\mnras] {10.1093/mnras/sty2448},
  \href {https://ui.adsabs.harvard.edu/abs/2018MNRAS.481.3278R} {481, 3278}

\bibitem[\protect\citeauthoryear{{Robertson} et~al.,}{{Robertson}
  et~al.}{2023}]{robertson2023}
{Robertson} B.~E.,  et~al., 2023, \mn@doi [\apjl] {10.3847/2041-8213/aca086},
  \href {https://ui.adsabs.harvard.edu/abs/2023ApJ...942L..42R} {942, L42}

\bibitem[\protect\citeauthoryear{{Rodighiero}, {Bisigello}, {Iani}, {Marasco},
  {Grazian}, {Sinigaglia}, {Cassata}  \& {Gruppioni}}{{Rodighiero}
  et~al.}{2023}]{rodighiero2023}
{Rodighiero} G.,  {Bisigello} L.,  {Iani} E.,  {Marasco} A.,  {Grazian} A.,
  {Sinigaglia} F.,  {Cassata} P.,   {Gruppioni} C.,  2023, \mn@doi [\mnras]
  {10.1093/mnrasl/slac115}, \href
  {https://ui.adsabs.harvard.edu/abs/2023MNRAS.518L..19R} {518, L19}

\bibitem[\protect\citeauthoryear{{Sanders}, {Shapley}, {Topping}, {Reddy}  \&
  {Brammer}}{{Sanders} et~al.}{2023a}]{sanders2023a}
{Sanders} R.~L.,  {Shapley} A.~E.,  {Topping} M.~W.,  {Reddy} N.~A.,
  {Brammer} G.~B.,  2023a, \mn@doi [arXiv e-prints]
  {10.48550/arXiv.2301.06696}, \href
  {https://ui.adsabs.harvard.edu/abs/2023arXiv230106696S} {p. arXiv:2301.06696}

\bibitem[\protect\citeauthoryear{{Sanders}, {Shapley}, {Topping}, {Reddy}  \&
  {Brammer}}{{Sanders} et~al.}{2023b}]{sanders2023b}
{Sanders} R.~L.,  {Shapley} A.~E.,  {Topping} M.~W.,  {Reddy} N.~A.,
  {Brammer} G.~B.,  2023b, \mn@doi [arXiv e-prints]
  {10.48550/arXiv.2303.08149}, \href
  {https://ui.adsabs.harvard.edu/abs/2023arXiv230308149S} {p. arXiv:2303.08149}

\bibitem[\protect\citeauthoryear{{Sassano}, {Schneider}, {Valiante},
  {Inayoshi}, {Chon}, {Omukai}, {Mayer}  \& {Capelo}}{{Sassano}
  et~al.}{2021}]{sassano2021}
{Sassano} F.,  {Schneider} R.,  {Valiante} R.,  {Inayoshi} K.,  {Chon} S.,
  {Omukai} K.,  {Mayer} L.,   {Capelo} P.~R.,  2021, \mn@doi [\mnras]
  {10.1093/mnras/stab1737}, \href
  {https://ui.adsabs.harvard.edu/abs/2021MNRAS.506..613S} {506, 613}

\bibitem[\protect\citeauthoryear{{Sassano}, {Capelo}, {Mayer}, {Schneider}  \&
  {Valiante}}{{Sassano} et~al.}{2023}]{sassano2023}
{Sassano} F.,  {Capelo} P.~R.,  {Mayer} L.,  {Schneider} R.,   {Valiante} R.,
  2023, \mn@doi [\mnras] {10.1093/mnras/stac3608}, \href
  {https://ui.adsabs.harvard.edu/abs/2023MNRAS.519.1837S} {519, 1837}

\bibitem[\protect\citeauthoryear{{Scholtz} et~al.,}{{Scholtz}
  et~al.}{2023}]{scholtz2023}
{Scholtz} J.,  et~al., 2023, \mn@doi [arXiv e-prints]
  {10.48550/arXiv.2306.09142}, \href
  {https://ui.adsabs.harvard.edu/abs/2023arXiv230609142S} {p. arXiv:2306.09142}

\bibitem[\protect\citeauthoryear{{Senchyna}, {Plat}, {Stark}  \&
  {Rudie}}{{Senchyna} et~al.}{2023}]{senchyna2023}
{Senchyna} P.,  {Plat} A.,  {Stark} D.~P.,   {Rudie} G.~C.,  2023, \mn@doi
  [arXiv e-prints] {10.48550/arXiv.2303.04179}, \href
  {https://ui.adsabs.harvard.edu/abs/2023arXiv230304179S} {p. arXiv:2303.04179}

\bibitem[\protect\citeauthoryear{{Smith}, {Regan}, {Downes}, {Norman}, {O'Shea}
   \& {Wise}}{{Smith} et~al.}{2018}]{smith2018}
{Smith} B.~D.,  {Regan} J.~A.,  {Downes} T.~P.,  {Norman} M.~L.,  {O'Shea}
  B.~W.,   {Wise} J.~H.,  2018, \mn@doi [\mnras] {10.1093/mnras/sty2103}, \href
  {https://ui.adsabs.harvard.edu/abs/2018MNRAS.480.3762S} {480, 3762}

\bibitem[\protect\citeauthoryear{{Spinoso}, {Bonoli}, {Valiante}, {Schneider}
  \& {Izquierdo-Villalba}}{{Spinoso} et~al.}{2022}]{spinoso2022}
{Spinoso} D.,  {Bonoli} S.,  {Valiante} R.,  {Schneider} R.,
  {Izquierdo-Villalba} D.,  2022, arXiv e-prints, \href
  {https://ui.adsabs.harvard.edu/abs/2022arXiv220313846S} {p. arXiv:2203.13846}

\bibitem[\protect\citeauthoryear{{Tacchella} et~al.,}{{Tacchella}
  et~al.}{2022}]{tacchella2022}
{Tacchella} S.,  et~al., 2022, \mn@doi [arXiv e-prints]
  {10.48550/arXiv.2208.03281}, \href
  {https://ui.adsabs.harvard.edu/abs/2022arXiv220803281T} {p. arXiv:2208.03281}

\bibitem[\protect\citeauthoryear{{Tacchella} et~al.,}{{Tacchella}
  et~al.}{2023}]{tacchella2023}
{Tacchella} S.,  et~al., 2023, \mn@doi [arXiv e-prints]
  {10.48550/arXiv.2302.07234}, \href
  {https://ui.adsabs.harvard.edu/abs/2023arXiv230207234T} {p. arXiv:2302.07234}

\bibitem[\protect\citeauthoryear{{Trinca}, {Schneider}, {Valiante}, {Graziani},
  {Zappacosta}  \& {Shankar}}{{Trinca} et~al.}{2022}]{trinca2022}
{Trinca} A.,  {Schneider} R.,  {Valiante} R.,  {Graziani} L.,  {Zappacosta} L.,
    {Shankar} F.,  2022, \mn@doi [\mnras] {10.1093/mnras/stac062}, \href
  {https://ui.adsabs.harvard.edu/abs/2022MNRAS.511..616T} {511, 616}

\bibitem[\protect\citeauthoryear{{Trinca}, {Schneider}, {Valiante}, {Graziani},
  {Ferrotti}, {Omukai}  \& {Chon}}{{Trinca} et~al.}{2023a}]{trinca2023gal}
{Trinca} A.,  {Schneider} R.,  {Valiante} R.,  {Graziani} L.,  {Ferrotti} A.,
  {Omukai} K.,   {Chon} S.,  2023a, \mn@doi [arXiv e-prints]
  {10.48550/arXiv.2305.04944}, \href
  {https://ui.adsabs.harvard.edu/abs/2023arXiv230504944T} {p. arXiv:2305.04944}

\bibitem[\protect\citeauthoryear{{Trinca}, {Schneider}, {Maiolino}, {Valiante},
  {Graziani}  \& {Volonteri}}{{Trinca} et~al.}{2023b}]{trinca2023bh}
{Trinca} A.,  {Schneider} R.,  {Maiolino} R.,  {Valiante} R.,  {Graziani} L.,
  {Volonteri} M.,  2023b, \mn@doi [\mnras] {10.1093/mnras/stac3768}, \href
  {https://ui.adsabs.harvard.edu/abs/2023MNRAS.519.4753T} {519, 4753}

\bibitem[\protect\citeauthoryear{{Trump} et~al.,}{{Trump}
  et~al.}{2023}]{trump2023}
{Trump} J.~R.,  et~al., 2023, \mn@doi [\apj] {10.3847/1538-4357/acba8a}, \href
  {https://ui.adsabs.harvard.edu/abs/2023ApJ...945...35T} {945, 35}

\bibitem[\protect\citeauthoryear{{Trussler}, {Conselice}, {Adams}, {Maiolino},
  {Nakajima}, {Zackrisson}  \& {Ferreira}}{{Trussler}
  et~al.}{2022}]{trussler2022}
{Trussler} J. A.~A.,  {Conselice} C.~J.,  {Adams} N.~J.,  {Maiolino} R.,
  {Nakajima} K.,  {Zackrisson} E.,   {Ferreira} L.,  2022, \mn@doi [arXiv
  e-prints] {10.48550/arXiv.2211.02038}, \href
  {https://ui.adsabs.harvard.edu/abs/2022arXiv221102038T} {p. arXiv:2211.02038}

\bibitem[\protect\citeauthoryear{{{\"U}bler} et~al.,}{{{\"U}bler}
  et~al.}{2023}]{ubler2023}
{{\"U}bler} H.,  et~al., 2023, \mn@doi [arXiv e-prints]
  {10.48550/arXiv.2302.06647}, \href
  {https://ui.adsabs.harvard.edu/abs/2023arXiv230206647U} {p. arXiv:2302.06647}

\bibitem[\protect\citeauthoryear{Valiante, Schneider, Salvadori  \&
  Gallerani}{Valiante et~al.}{2014}]{valiante2014}
Valiante R.,  Schneider R.,  Salvadori S.,   Gallerani S.,  2014, Monthly
  Notices of the Royal Astronomical Society, 444, 2442

\bibitem[\protect\citeauthoryear{Valiante, Schneider, Volonteri  \&
  Omukai}{Valiante et~al.}{2016}]{valiante2016}
Valiante R.,  Schneider R.,  Volonteri M.,   Omukai K.,  2016, Monthly Notices
  of the Royal Astronomical Society, 457, 3356

\bibitem[\protect\citeauthoryear{{Valiante}, {Schneider}, {Zappacosta},
  {Graziani}, {Pezzulli}  \& {Volonteri}}{{Valiante}
  et~al.}{2018}]{valiante2018observability}
{Valiante} R.,  {Schneider} R.,  {Zappacosta} L.,  {Graziani} L.,  {Pezzulli}
  E.,   {Volonteri} M.,  2018, \mn@doi [\mnras] {10.1093/mnras/sty213}, \href
  {https://ui.adsabs.harvard.edu/abs/2018MNRAS.476..407V} {476, 407}

\bibitem[\protect\citeauthoryear{{Ventura}, {Trinca}, {Schneider}, {Graziani},
  {Valiante}  \& {Wyithe}}{{Ventura} et~al.}{2023}]{ventura2023}
{Ventura} E.~M.,  {Trinca} A.,  {Schneider} R.,  {Graziani} L.,  {Valiante} R.,
    {Wyithe} J. S.~B.,  2023, \mn@doi [\mnras] {10.1093/mnras/stad237}, \href
  {https://ui.adsabs.harvard.edu/abs/2023MNRAS.tmp..252V} {}

\bibitem[\protect\citeauthoryear{{Visbal} \& {Haiman}}{{Visbal} \&
  {Haiman}}{2018}]{visbal2018}
{Visbal} E.,  {Haiman} Z.,  2018, \mn@doi [\apjl] {10.3847/2041-8213/aadf3a},
  \href {https://ui.adsabs.harvard.edu/abs/2018ApJ...865L...9V} {865, L9}

\bibitem[\protect\citeauthoryear{{Volonteri}}{{Volonteri}}{2012}]{volonteri2012}
{Volonteri} M.,  2012, \mn@doi [Science] {10.1126/science.1220843}, \href
  {https://ui.adsabs.harvard.edu/abs/2012Sci...337..544V} {337, 544}

\bibitem[\protect\citeauthoryear{{Volonteri} \& {Natarajan}}{{Volonteri} \&
  {Natarajan}}{2009}]{volonteri2009}
{Volonteri} M.,  {Natarajan} P.,  2009, \mn@doi [\mnras]
  {10.1111/j.1365-2966.2009.15577.x}, \href
  {https://ui.adsabs.harvard.edu/abs/2009MNRAS.400.1911V} {400, 1911}

\bibitem[\protect\citeauthoryear{{Volonteri} \& {Stark}}{{Volonteri} \&
  {Stark}}{2011}]{volonteri2011}
{Volonteri} M.,  {Stark} D.~P.,  2011, \mn@doi [\mnras]
  {10.1111/j.1365-2966.2011.19391.x}, \href
  {https://ui.adsabs.harvard.edu/abs/2011MNRAS.417.2085V} {417, 2085}

\bibitem[\protect\citeauthoryear{{Volonteri}, {Silk}  \& {Dubus}}{{Volonteri}
  et~al.}{2015}]{volonteri2015}
{Volonteri} M.,  {Silk} J.,   {Dubus} G.,  2015, \mn@doi [\apj]
  {10.1088/0004-637X/804/2/148}, \href
  {https://ui.adsabs.harvard.edu/abs/2015ApJ...804..148V} {804, 148}

\bibitem[\protect\citeauthoryear{{Volonteri}, {Reines}, {Atek}, {Stark}  \&
  {Trebitsch}}{{Volonteri} et~al.}{2017}]{volonteri2017}
{Volonteri} M.,  {Reines} A.~E.,  {Atek} H.,  {Stark} D.~P.,   {Trebitsch} M.,
  2017, \mn@doi [\apj] {10.3847/1538-4357/aa93f1}, \href
  {https://ui.adsabs.harvard.edu/abs/2017ApJ...849..155V} {849, 155}

\bibitem[\protect\citeauthoryear{{Volonteri}, {Habouzit}  \&
  {Colpi}}{{Volonteri} et~al.}{2021}]{volonteri2021}
{Volonteri} M.,  {Habouzit} M.,   {Colpi} M.,  2021, \mn@doi [Nature Reviews
  Physics] {10.1038/s42254-021-00364-9}, \href
  {https://ui.adsabs.harvard.edu/abs/2021NatRP...3..732V} {3, 732}

\bibitem[\protect\citeauthoryear{{Volonteri}, {Habouzit}  \&
  {Colpi}}{{Volonteri} et~al.}{2023}]{volonteri2023}
{Volonteri} M.,  {Habouzit} M.,   {Colpi} M.,  2023, \mn@doi [\mnras]
  {10.1093/mnras/stad499}, \href
  {https://ui.adsabs.harvard.edu/abs/2023MNRAS.521..241V} {521, 241}

\bibitem[\protect\citeauthoryear{{Wang} et~al.,}{{Wang}
  et~al.}{2021}]{wang2021}
{Wang} F.,  et~al., 2021, \mn@doi [\apjl] {10.3847/2041-8213/abd8c6}, \href
  {https://ui.adsabs.harvard.edu/abs/2021ApJ...907L...1W} {907, L1}

\bibitem[\protect\citeauthoryear{{Whitler}, {Stark}, {Endsley}, {Leja},
  {Charlot}  \& {Chevallard}}{{Whitler} et~al.}{2023}]{whitler2023}
{Whitler} L.,  {Stark} D.~P.,  {Endsley} R.,  {Leja} J.,  {Charlot} S.,
  {Chevallard} J.,  2023, \mn@doi [\mnras] {10.1093/mnras/stad004}, \href
  {https://ui.adsabs.harvard.edu/abs/2023MNRAS.519.5859W} {519, 5859}

\bibitem[\protect\citeauthoryear{{Yang} et~al.,}{{Yang}
  et~al.}{2020}]{yang2020}
{Yang} J.,  et~al., 2020, \mn@doi [\apjl] {10.3847/2041-8213/ab9c26}, \href
  {https://ui.adsabs.harvard.edu/abs/2020ApJ...897L..14Y} {897, L14}

\bibitem[\protect\citeauthoryear{{Zavala} et~al.,}{{Zavala}
  et~al.}{2023}]{zavala2023}
{Zavala} J.~A.,  et~al., 2023, \mn@doi [\apjl] {10.3847/2041-8213/acacfe},
  \href {https://ui.adsabs.harvard.edu/abs/2023ApJ...943L...9Z} {943, L9}

\makeatother
\end{thebibliography}


\bsp	
\label{lastpage}
\end{document}